\ifx\mnmacrosloaded\undefined 

%
%
%
%

\catcode `\@=11 

\def\@version{1.4}
\def\@verdate{22nd Feb 1994}

%
%
%
%


\newif\ifprod@font

\ifx\@typeface\undefined
  \def\@typeface{Comp. Modern}\prod@fontfalse
\else
  \prod@fonttrue 
\fi

\def\newfam{\alloc@8\fam\chardef\sixt@@n} 

\ifprod@font
\font\fiverm=mtr10 at 5pt
\font\fivebf=mtbx10 at 5pt
\font\fiveit=mtti10 at 5pt
\font\fivesl=mtsl10 at 5pt
\font\fivett=mttt10 at 5pt     \hyphenchar\fivett=-1
\font\fivecsc=mtcsc10 at 5pt
\font\fivesf=mtss10 at 5pt
\font\fivei=mtmi10 at 5pt      \skewchar\fivei='177
\font\fivemib=mtmib10 at 5pt   \skewchar\fivemib='177
\font\fivesy=mtsy10 at 5pt     \skewchar\fivesy='60
\font\fivesyb=mtbsy10 at 5pt   \skewchar\fivesyb='60

\font\sixrm=mtr10 at 6pt
\font\sixbf=mtbx10 at 6pt
\font\sixit=mtti10 at 6pt
\font\sixsl=mtsl10 at 6pt
\font\sixtt=mttt10 at 6pt      \hyphenchar\sixtt=-1
\font\sixcsc=mtcsc10 at 6pt
\font\sixsf=mtss10 at 6pt
\font\sixi=mtmi10 at 6pt       \skewchar\sixi='177
\font\sixmib=mtmib10 at 6pt    \skewchar\sixmib='177
\font\sixsy=mtsy10 at 6pt      \skewchar\sixsy='60
\font\sixsyb=mtbsy10 at 6pt    \skewchar\sixsyb='60

\font\sevenrm=mtr10 at 7pt
\font\sevenbf=mtbx10 at 7pt
\font\sevenit=mtti10 at 7pt
\font\sevensl=mtsl10 at 7pt
\font\seventt=mttt10 at 7pt     \hyphenchar\seventt=-1
\font\sevencsc=mtcsc10 at 7pt
\font\sevensf=mtss10 at 7pt
\font\seveni=mtmi10 at 7pt      \skewchar\seveni='177
\font\sevenmib=mtmib10 at 7pt   \skewchar\sevenmib='177
\font\sevensy=mtsy10 at 7pt     \skewchar\sevensy='60
\font\sevensyb=mtbsy10 at 7pt   \skewchar\sevensyb='60

\font\eightrm=mtr10 at 8pt
\font\eightbf=mtbx10 at 8pt
\font\eightit=mtti10 at 8pt
\font\eighti=mtmi10 at 8pt      \skewchar\eighti='177
\font\eightmib=mtmib10 at 8pt   \skewchar\eightmib='177
\font\eightsy=mtsy10 at 8pt     \skewchar\eightsy='60
\font\eightsyb=mtbsy10 at 8pt   \skewchar\eightsyb='60
\font\eightsl=mtsl10 at 8pt
\font\eighttt=mttt10 at 8pt     \hyphenchar\eighttt=-1
\font\eightcsc=mtcsc10 at 8pt
\font\eightsf=mtss10 at 8pt

\font\ninerm=mtr10 at 9pt
\font\ninebf=mtbx10 at 9pt
\font\nineit=mtti10 at 9pt
\font\ninei=mtmi10 at 9pt      \skewchar\ninei='177
\font\ninemib=mtmib10 at 9pt   \skewchar\ninemib='177
\font\ninesy=mtsy10 at 9pt     \skewchar\ninesy='60
\font\ninesyb=mtbsy10 at 9pt   \skewchar\ninesyb='60
\font\ninesl=mtsl10 at 9pt
\font\ninett=mttt10 at 9pt     \hyphenchar\ninett=-1
\font\ninecsc=mtcsc10 at 9pt
\font\ninesf=mtss10 at 9pt

\font\tenrm=mtr10
\font\tenbf=mtbx10
\font\tenit=mtti10
\font\teni=mtmi10		\skewchar\teni='177
\font\tenmib=mtmib10	\skewchar\tenmib='177
\font\tensy=mtsy10		\skewchar\tensy='60
\font\tensyb=mtbsy10	\skewchar\tensyb='60
\font\tenex=cmex10
\font\tensl=mtsl10
\font\tentt=mttt10		\hyphenchar\tentt=-1
\font\tencsc=mtcsc10
\font\tensf=mtss10

\font\elevenrm=mtr10 at 11pt
\font\elevenbf=mtbx10 at 11pt
\font\elevenit=mtti10 at 11pt
\font\eleveni=mtmi10 at 11pt      \skewchar\eleveni='177
\font\elevenmib=mtmib10 at 11pt   \skewchar\elevenmib='177
\font\elevensy=mtsy10 at 11pt     \skewchar\elevensy='60
\font\elevensyb=mtbsy10 at 11pt   \skewchar\elevensyb='60
\font\elevensl=mtsl10 at 11pt
\font\eleventt=mttt10 at 11pt     \hyphenchar\eleventt=-1
\font\elevencsc=mtcsc10 at 11pt
\font\elevensf=mtss10 at 11pt

\font\twelverm=mtr10 at 12pt
\font\twelvebf=mtbx10 at 12pt
\font\twelveit=mtti10 at 12pt
\font\twelvesl=mtsl10 at 12pt
\font\twelvett=mttt10 at 12pt     \hyphenchar\twelvett=-1
\font\twelvecsc=mtcsc10 at 12pt
\font\twelvesf=mtss10 at 12pt
\font\twelvei=mtmi10 at 12pt      \skewchar\twelvei='177
\font\twelvemib=mtmib10 at 12pt   \skewchar\twelvemib='177
\font\twelvesy=mtsy10 at 12pt     \skewchar\twelvesy='60
\font\twelvesyb=mtbsy10 at 12pt   \skewchar\twelvesyb='60

\font\fourteenrm=mtr10 at 14pt
\font\fourteenbf=mtbx10 at 14pt
\font\fourteenit=mtti10 at 14pt
\font\fourteeni=mtmi10 at 14pt      \skewchar\fourteeni='177
\font\fourteenmib=mtmib10 at 14pt   \skewchar\fourteenmib='177
\font\fourteensy=mtsy10 at 14pt     \skewchar\fourteensy='60
\font\fourteensyb=mtbsy10 at 14pt   \skewchar\fourteensyb='60
\font\fourteensl=mtsl10 at 14pt
\font\fourteentt=mttt10 at 14pt     \hyphenchar\fourteentt=-1
\font\fourteencsc=mtcsc10 at 14pt
\font\fourteensf=mtss10 at 14pt

\font\seventeenrm=mtr10 at 17pt
\font\seventeenbf=mtbx10 at 17pt
\font\seventeenit=mtti10 at 17pt
\font\seventeeni=mtmi10 at 17pt      \skewchar\seventeeni='177
\font\seventeenmib=mtmib10 at 17pt   \skewchar\seventeenmib='177
\font\seventeensy=mtsy10 at 17pt     \skewchar\seventeensy='60
\font\seventeensyb=mtbsy10 at 17pt   \skewchar\seventeensyb='60
\font\seventeensl=mtsl10 at 17pt
\font\seventeentt=mttt10 at 17pt     \hyphenchar\seventeentt=-1
\font\seventeencsc=mtcsc10 at 17pt
\font\seventeensf=mtss10 at 17pt


\newfam\xmfam
\newfam\ymfam

\font\fivexm=mtxm10 at 5pt
\font\sixxm=mtxm10 at 6pt
\font\sevenxm=mtxm10 at 7pt
\font\eightxm=mtxm10 at 8pt
\font\ninexm=mtxm10 at 9pt
\font\tenxm=mtxm10
\font\elevenxm=mtxm10 at 11pt
\font\twelvexm=mtxm10 at 12pt
\font\fourteenxm=mtxm10 at 14pt
\font\seventeenxm=mtxm10 at 17pt

\font\fiveym=mtym10 at 5pt
\font\sixym=mtym10 at 6pt
\font\sevenym=mtym10 at 7pt
\font\eightym=mtym10 at 8pt
\font\nineym=mtym10 at 9pt
\font\tenym=mtym10
\font\elevenym=mtym10 at 11pt
\font\twelveym=mtym10 at 12pt
\font\fourteenym=mtym10 at 14pt
\font\seventeenym=mtym10 at 17pt
\else
\font\fiverm=cmr5
\font\fivei=cmmi5             \skewchar\fivei='177
\font\fivemib=cmmib10 at 5pt  \skewchar\fivemib='177
\font\fivesy=cmsy5            \skewchar\fivesy='60
\font\fivesyb=cmbsy10 at 5pt  \skewchar\fivesyb='60
\font\fivebf=cmbx5

\font\sixrm=cmr6
\font\sixi=cmmi6             \skewchar\sixi='177
\font\sixmib=cmmib10 at 6pt  \skewchar\sixmib='177
\font\sixsy=cmsy6            \skewchar\sixsy='60
\font\sixsyb=cmbsy10 at 6pt  \skewchar\sixsyb='60
\font\sixbf=cmbx6

\font\sevenrm=cmr7
\font\seveni=cmmi7             \skewchar\seveni='177
\font\sevenmib=cmmib10 at 7pt  \skewchar\sevenmib='177
\font\sevensy=cmsy7            \skewchar\sevensy='60
\font\sevensyb=cmbsy10 at 7pt  \skewchar\sevensyb='60
\font\sevenbf=cmbx7

\font\eightrm=cmr8
\font\eightbf=cmbx8
\font\eightit=cmti8
\font\eighti=cmmi8			\skewchar\eighti='177
\font\eightmib=cmmib10 at 8pt	\skewchar\eightmib='177
\font\eightsy=cmsy8			\skewchar\eightsy='60
\font\eightsyb=cmbsy10 at 8pt	\skewchar\eightsyb='60
\font\eightsl=cmsl8
\font\eighttt=cmtt8			\hyphenchar\eighttt=-1
\font\eightcsc=cmcsc10 at 8pt
\font\eightsf=cmss8

\font\ninerm=cmr9
\font\ninebf=cmbx9
\font\nineit=cmti9
\font\ninei=cmmi9			\skewchar\ninei='177
\font\ninemib=cmmib10 at 9pt	\skewchar\ninemib='177
\font\ninesy=cmsy9			\skewchar\ninesy='60
\font\ninesyb=cmbsy10 at 9pt	\skewchar\ninesyb='60
\font\ninesl=cmsl9
\font\ninett=cmtt9			\hyphenchar\ninett=-1
\font\ninecsc=cmcsc10 at 9pt
\font\ninesf=cmss9

\font\tenrm=cmr10
\font\tenbf=cmbx10
\font\tenit=cmti10
\font\teni=cmmi10		\skewchar\teni='177
\font\tenmib=cmmib10	\skewchar\tenmib='177
\font\tensy=cmsy10		\skewchar\tensy='60
\font\tensyb=cmbsy10	\skewchar\tensyb='60
\font\tenex=cmex10
\font\tensl=cmsl10
\font\tentt=cmtt10		\hyphenchar\tentt=-1
\font\tencsc=cmcsc10
\font\tensf=cmss10

\font\elevenrm=cmr10 scaled \magstephalf
\font\elevenbf=cmbx10 scaled \magstephalf
\font\elevenit=cmti10 scaled \magstephalf
\font\eleveni=cmmi10 scaled \magstephalf	\skewchar\eleveni='177
\font\elevenmib=cmmib10 scaled \magstephalf	\skewchar\elevenmib='177
\font\elevensy=cmsy10 scaled \magstephalf	\skewchar\elevensy='60
\font\elevensyb=cmbsy10 scaled \magstephalf	\skewchar\elevensyb='60
\font\elevensl=cmsl10 scaled \magstephalf
\font\eleventt=cmtt10 scaled \magstephalf	\hyphenchar\eleventt=-1
\font\elevencsc=cmcsc10 scaled \magstephalf
\font\elevensf=cmss10 scaled \magstephalf

\font\twelverm=cmr10 scaled \magstep1
\font\twelvebf=cmbx10 scaled \magstep1
\font\twelvei=cmmi10 scaled \magstep1      \skewchar\twelvei='177
\font\twelvemib=cmmib10 scaled \magstep1   \skewchar\twelvemib='177
\font\twelvesy=cmsy10 scaled \magstep1     \skewchar\twelvesy='60
\font\twelvesyb=cmbsy10 scaled \magstep1   \skewchar\twelvesyb='60

\font\fourteenrm=cmr10 scaled \magstep2
\font\fourteenbf=cmbx10 scaled \magstep2
\font\fourteenit=cmti10 scaled \magstep2
\font\fourteeni=cmmi10 scaled \magstep2		\skewchar\fourteeni='177
\font\fourteenmib=cmmib10 scaled \magstep2	\skewchar\fourteenmib='177
\font\fourteensy=cmsy10 scaled \magstep2	\skewchar\fourteensy='60
\font\fourteensyb=cmbsy10 scaled \magstep2	\skewchar\fourteensyb='60
\font\fourteensl=cmsl10 scaled \magstep2
\font\fourteentt=cmtt10 scaled \magstep2	\hyphenchar\fourteentt=-1
\font\fourteencsc=cmcsc10 scaled \magstep2
\font\fourteensf=cmss10 scaled \magstep2

\font\seventeenrm=cmr10 scaled \magstep3
\font\seventeenbf=cmbx10 scaled \magstep3
\font\seventeenit=cmti10 scaled \magstep3
\font\seventeeni=cmmi10 scaled \magstep3	\skewchar\seventeeni='177
\font\seventeenmib=cmmib10 scaled \magstep3	\skewchar\seventeenmib='177
\font\seventeensy=cmsy10 scaled \magstep3	\skewchar\seventeensy='60
\font\seventeensyb=cmbsy10 scaled \magstep3	\skewchar\seventeensyb='60
\font\seventeensl=cmsl10 scaled \magstep3
\font\seventeentt=cmtt10 scaled \magstep3	\hyphenchar\seventeentt=-1
\font\seventeencsc=cmcsc10 scaled \magstep3
\font\seventeensf=cmss10 scaled \magstep3
\fi

\def\hexnumber#1{\ifcase#1 0\or1\or2\or3\or4\or5\or6\or7\or8\or9\or
  A\or B\or C\or D\or E\or F\fi}

\ifprod@font
  \edef\@xm{\hexnumber\xmfam}
  \edef\@ym{\hexnumber\ymfam}
\fi

\def\makestrut{%
  \setbox\strutbox=\hbox{%
    \vrule height.7\baselineskip depth.3\baselineskip width \z@}%
}

\def\baselinestretch{1}
\newskip\tmp@bls

\def\b@ls#1{
  \tmp@bls=#1\relax
  \baselineskip=#1\relax\makestrut
  \normalbaselineskip=\baselinestretch\tmp@bls
  \normalbaselines
}

\def\nostb@ls#1{
  \normalbaselineskip=#1\relax
  \normalbaselines
  \makestrut
}

%

\newfam\mibfam 
\newfam\sybfam 
\newfam\scfam  
\newfam\sffam  

\def\mit{\fam\@ne}

\def\cal{\fam\tw@}

\def\em{\ifdim\fontdimen1\font>\z@ \rm\else\it\fi}

\textfont3=\tenex
\scriptfont3=\tenex
\scriptscriptfont3=\tenex

\setbox0=\hbox{\tenex B} \p@renwd=\wd0 

\def\eightpoint{
  \def\rm{\fam0\eightrm}%
  \textfont0=\eightrm \scriptfont0=\sixrm \scriptscriptfont0=\fiverm%
  \textfont1=\eighti  \scriptfont1=\sixi  \scriptscriptfont1=\fivei%
  \textfont2=\eightsy \scriptfont2=\sixsy \scriptscriptfont2=\fivesy%
  \textfont\itfam=\eightit\def\it{\fam\itfam\eightit}%
  \ifprod@font
    \scriptfont\itfam=\sixit
      \scriptscriptfont\itfam=\fiveit
  \else
    \scriptfont\itfam=\eightit
      \scriptscriptfont\itfam=\eightit
  \fi
  \textfont\bffam=\eightbf%
    \scriptfont\bffam=\sixbf%
      \scriptscriptfont\bffam=\fivebf%
  \def\bf{\fam\bffam\eightbf}%
  \textfont\slfam=\eightsl\def\sl{\fam\slfam\eightsl}%
  \ifprod@font
    \scriptfont\slfam=\sixsl
      \scriptscriptfont\slfam=\fivesl
  \else
    \scriptfont\slfam=\eightsl
      \scriptscriptfont\slfam=\eightsl
  \fi
  \textfont\ttfam=\eighttt\def\tt{\fam\ttfam\eighttt}%
  \ifprod@font
    \scriptfont\ttfam=\sixtt
      \scriptscriptfont\ttfam=\fivett
  \else
    \scriptfont\ttfam=\eighttt
      \scriptscriptfont\ttfam=\eighttt
  \fi
  \textfont\scfam=\eightcsc\def\sc{\fam\scfam\eightcsc}%
  \ifprod@font
    \scriptfont\scfam=\sixcsc
      \scriptscriptfont\scfam=\fivecsc
  \else
    \scriptfont\scfam=\eightcsc
      \scriptscriptfont\scfam=\eightcsc
  \fi
  \textfont\sffam=\eightsf\def\sf{\fam\sffam\eightsf}%
  \ifprod@font
    \scriptfont\sffam=\sixsf
      \scriptscriptfont\sffam=\fivesf
  \else
    \scriptfont\sffam=\eightsf
      \scriptscriptfont\sffam=\eightsf
  \fi
  \textfont\mibfam=\eightmib
    \scriptfont\mibfam=\sixmib
      \scriptscriptfont\mibfam=\fivemib
  \textfont\sybfam=\eightsyb
    \scriptfont\sybfam=\sixsyb
      \scriptscriptfont\sybfam=\fivesyb
  \ifprod@font
    \textfont\xmfam=\eightxm
      \scriptfont\xmfam=\sixxm
        \scriptscriptfont\xmfam=\fivexm
    \textfont\ymfam=\eightym
      \scriptfont\ymfam=\sixym
        \scriptscriptfont\ymfam=\fiveym
  \fi
  \def\oldstyle{\fam\@ne\eighti}%
  \def\boldstyle{\fam\mibfam\eightmib}%
  \b@ls{10pt}\rm%
}

\def\ninepoint{
  \def\rm{\fam0\ninerm}%
  \textfont0=\ninerm \scriptfont0=\sixrm \scriptscriptfont0=\fiverm%
  \textfont1=\ninei  \scriptfont1=\sixi  \scriptscriptfont1=\fivei%
  \textfont2=\ninesy \scriptfont2=\sixsy \scriptscriptfont2=\fivesy%
  \textfont\itfam=\nineit\def\it{\fam\itfam\nineit}%
  \ifprod@font
    \scriptfont\itfam=\sixit
      \scriptscriptfont\itfam=\fiveit
  \else
    \scriptfont\itfam=\nineit
      \scriptscriptfont\itfam=\nineit
  \fi
  \textfont\bffam=\ninebf%
    \scriptfont\bffam=\sixbf%
      \scriptscriptfont\bffam=\fivebf%
  \def\bf{\fam\bffam\ninebf}%
  \textfont\slfam=\ninesl\def\sl{\fam\slfam\ninesl}%
  \ifprod@font
    \scriptfont\slfam=\sixsl
      \scriptscriptfont\slfam=\fivesl
  \else
    \scriptfont\slfam=\ninesl
      \scriptscriptfont\slfam=\ninesl
  \fi
  \textfont\ttfam=\ninett\def\tt{\fam\ttfam\ninett}%
  \ifprod@font
    \scriptfont\ttfam=\sixtt
      \scriptscriptfont\ttfam=\fivett
  \else
    \scriptfont\ttfam=\ninett
      \scriptscriptfont\ttfam=\ninett
  \fi
  \textfont\scfam=\ninecsc\def\sc{\fam\scfam\ninecsc}%
  \ifprod@font
    \scriptfont\scfam=\sixcsc
      \scriptscriptfont\scfam=\fivecsc
  \else
    \scriptfont\scfam=\ninecsc
      \scriptscriptfont\scfam=\ninecsc
  \fi
  \textfont\sffam=\ninesf\def\sf{\fam\sffam\ninesf}%
  \ifprod@font
    \scriptfont\sffam=\sixsf
      \scriptscriptfont\sffam=\fivesf
  \else
    \scriptfont\sffam=\ninesf
      \scriptscriptfont\sffam=\ninesf
  \fi
  \textfont\mibfam=\ninemib
    \scriptfont\mibfam=\sixmib
      \scriptscriptfont\mibfam=\fivemib
  \textfont\sybfam=\ninesyb
    \scriptfont\sybfam=\sixsyb
      \scriptscriptfont\sybfam=\fivesyb
  \ifprod@font
    \textfont\xmfam=\ninexm
      \scriptfont\xmfam=\sixxm
        \scriptscriptfont\xmfam=\fivexm
    \textfont\ymfam=\nineym
      \scriptfont\ymfam=\sixym
        \scriptscriptfont\ymfam=\fiveym
  \fi
  \def\oldstyle{\fam\@ne\ninei}%
  \def\boldstyle{\fam\mibfam\ninemib}%
  \b@ls{\TextLeading plus \Feathering}\rm%
}

\def\tenpoint{
  \def\rm{\fam0\tenrm}%
  \textfont0=\tenrm \scriptfont0=\sevenrm \scriptscriptfont0=\fiverm%
  \textfont1=\teni  \scriptfont1=\seveni  \scriptscriptfont1=\fivei%
  \textfont2=\tensy \scriptfont2=\sevensy \scriptscriptfont2=\fivesy%
  \textfont\itfam=\tenit\def\it{\fam\itfam\tenit}%
  \ifprod@font
    \scriptfont\itfam=\sevenit
      \scriptscriptfont\itfam=\fiveit
  \else
    \scriptfont\itfam=\tenit
      \scriptscriptfont\itfam=\tenit
  \fi
  \textfont\bffam=\tenbf%
    \scriptfont\bffam=\sevenbf%
      \scriptscriptfont\bffam=\fivebf%
  \def\bf{\fam\bffam\tenbf}%
  \textfont\slfam=\tensl\def\sl{\fam\slfam\tensl}%
  \ifprod@font
    \scriptfont\slfam=\sevensl
      \scriptscriptfont\slfam=\fivesl
  \else
    \scriptfont\slfam=\tensl
      \scriptscriptfont\slfam=\tensl
  \fi
  \textfont\ttfam=\tentt\def\tt{\fam\ttfam\tentt}%
  \ifprod@font
    \scriptfont\ttfam=\seventt
      \scriptscriptfont\ttfam=\fivett
  \else
    \scriptfont\ttfam=\tentt
      \scriptscriptfont\ttfam=\tentt
  \fi
  \textfont\scfam=\tencsc\def\sc{\fam\scfam\tencsc}%
  \ifprod@font
    \scriptfont\scfam=\sevencsc
      \scriptscriptfont\scfam=\fivecsc
  \else
    \scriptfont\scfam=\tencsc
      \scriptscriptfont\scfam=\tencsc
  \fi
  \textfont\sffam=\tensf\def\sf{\fam\sffam\tensf}%
  \ifprod@font
    \scriptfont\sffam=\sevensf
      \scriptscriptfont\sffam=\fivesf
  \else
    \scriptfont\sffam=\tensf
      \scriptscriptfont\sffam=\tensf
  \fi
  \textfont\mibfam=\tenmib
    \scriptfont\mibfam=\sevenmib
      \scriptscriptfont\mibfam=\fivemib
  \textfont\sybfam=\tensyb
    \scriptfont\sybfam=\sevensyb
      \scriptscriptfont\sybfam=\fivesyb
  \ifprod@font
    \textfont\xmfam=\tenxm
      \scriptfont\xmfam=\sevenxm
        \scriptscriptfont\xmfam=\fivexm
    \textfont\ymfam=\tenym
      \scriptfont\ymfam=\sevenym
        \scriptscriptfont\ymfam=\fiveym
  \fi
  \def\oldstyle{\fam\@ne\teni}%
  \def\boldstyle{\fam\mibfam\tenmib}%
  \b@ls{11pt}\rm%
}

\def\elevenpoint{
  \def\rm{\fam0\elevenrm}%
  \textfont0=\elevenrm \scriptfont0=\eightrm \scriptscriptfont0=\sixrm%
  \textfont1=\eleveni  \scriptfont1=\eighti  \scriptscriptfont1=\sixi%
  \textfont2=\elevensy \scriptfont2=\eightsy \scriptscriptfont2=\sixsy%
  \textfont\itfam=\elevenit\def\it{\fam\itfam\elevenit}%
  \ifprod@font
    \scriptfont\itfam=\eightit
      \scriptscriptfont\itfam=\sixit
  \else
    \scriptfont\itfam=\elevenit
      \scriptscriptfont\itfam=\elevenit
  \fi
  \textfont\bffam=\elevenbf%
    \scriptfont\bffam=\eightbf%
      \scriptscriptfont\bffam=\sixbf%
  \def\bf{\fam\bffam\elevenbf}%
  \textfont\slfam=\elevensl\def\sl{\fam\slfam\elevensl}%
  \ifprod@font
    \scriptfont\slfam=\eightsl
      \scriptscriptfont\slfam=\sixsl
  \else
    \scriptfont\slfam=\elevensl
      \scriptscriptfont\slfam=\elevensl
  \fi
  \textfont\ttfam=\eleventt\def\tt{\fam\ttfam\eleventt}%
  \ifprod@font
    \scriptfont\ttfam=\eighttt
      \scriptscriptfont\ttfam=\sixtt
  \else
    \scriptfont\ttfam=\eleventt
      \scriptscriptfont\ttfam=\eleventt
  \fi
  \textfont\scfam=\elevencsc\def\sc{\fam\scfam\elevencsc}%
  \ifprod@font
    \scriptfont\scfam=\eightcsc
      \scriptscriptfont\scfam=\sixcsc
  \else
    \scriptfont\scfam=\elevencsc
      \scriptscriptfont\scfam=\elevencsc
  \fi
  \textfont\sffam=\elevensf\def\sf{\fam\sffam\elevensf}%
  \ifprod@font
    \scriptfont\sffam=\eightsf
      \scriptscriptfont\sffam=\sixsf
  \else
    \scriptfont\sffam=\elevensf
      \scriptscriptfont\sffam=\elevensf
  \fi
  \textfont\mibfam=\elevenmib
    \scriptfont\mibfam=\eightmib
      \scriptscriptfont\mibfam=\sixmib
  \textfont\sybfam=\elevensyb
    \scriptfont\sybfam=\eightsyb
      \scriptscriptfont\sybfam=\sixsyb
  \ifprod@font
    \textfont\xmfam=\elevenxm
      \scriptfont\xmfam=\eightxm
       \scriptscriptfont\xmfam=\sixxm
    \textfont\ymfam=\elevenym
      \scriptfont\ymfam=\eightym
        \scriptscriptfont\ymfam=\sixym
   \fi
  \def\oldstyle{\fam\@ne\eleveni}%
  \def\boldstyle{\fam\mibfam\elevenmib}%
  \b@ls{13pt}\rm%
}

\def\fourteenpoint{
  \def\rm{\fam0\fourteenrm}%
  \textfont0\fourteenrm  \scriptfont0\tenrm  \scriptscriptfont0\sevenrm%
  \textfont1\fourteeni   \scriptfont1\teni   \scriptscriptfont1\seveni%
  \textfont2\fourteensy  \scriptfont2\tensy  \scriptscriptfont2\sevensy%
  \textfont\itfam=\fourteenit\def\it{\fam\itfam\fourteenit}%
  \ifprod@font
    \scriptfont\itfam=\tenit
      \scriptscriptfont\itfam=\sevenit
  \else
    \scriptfont\itfam=\fourteenit
      \scriptscriptfont\itfam=\fourteenit
  \fi
  \textfont\bffam=\fourteenbf%
    \scriptfont\bffam=\tenbf%
      \scriptscriptfont\bffam=\sevenbf%
  \def\bf{\fam\bffam\fourteenbf}%
  \textfont\slfam=\fourteensl\def\sl{\fam\slfam\fourteensl}%
  \ifprod@font
    \scriptfont\slfam=\tensl
      \scriptscriptfont\slfam=\sevensl
  \else
    \scriptfont\slfam=\fourteensl
      \scriptscriptfont\slfam=\fourteensl
  \fi
  \textfont\ttfam=\fourteentt\def\tt{\fam\ttfam\fourteentt}%
  \ifprod@font
    \scriptfont\ttfam=\tentt
      \scriptscriptfont\ttfam=\seventt
  \else
    \scriptfont\ttfam=\fourteentt
      \scriptscriptfont\ttfam=\fourteentt
  \fi
  \textfont\scfam=\fourteencsc\def\sc{\fam\scfam\fourteencsc}%
  \ifprod@font
    \scriptfont\scfam=\tencsc
      \scriptscriptfont\scfam=\sevencsc
  \else
    \scriptfont\scfam=\fourteencsc
      \scriptscriptfont\scfam=\fourteencsc
  \fi
  \textfont\sffam=\fourteensf\def\sf{\fam\sffam\fourteensf}%
  \ifprod@font
    \scriptfont\sffam=\tensf
      \scriptscriptfont\sffam=\sevensf
  \else
    \scriptfont\sffam=\fourteensf
      \scriptscriptfont\sffam=\fourteensf
  \fi
  \textfont\mibfam=\fourteenmib
    \scriptfont\mibfam=\tenmib
      \scriptscriptfont\mibfam=\sevenmib
  \textfont\sybfam=\fourteensyb
    \scriptfont\sybfam=\tensyb
      \scriptscriptfont\sybfam=\sevensyb
  \ifprod@font
    \textfont\xmfam=\fourteenxm
      \scriptfont\xmfam=\tenxm
        \scriptscriptfont\xmfam=\sevenxm
   \textfont\ymfam=\fourteenym
      \scriptfont\ymfam=\tenym
        \scriptscriptfont\ymfam=\sevenym
  \fi
  \def\oldstyle{\fam\@ne\fourteeni}%
  \def\boldstyle{\fam\mibfam\fourteenmib}%
  \b@ls{17pt}\rm%
}

\def\seventeenpoint{
  \def\rm{\fam0\seventeenrm}%
  \textfont0\seventeenrm  \scriptfont0\twelverm  \scriptscriptfont0\tenrm%
  \textfont1\seventeeni   \scriptfont1\twelvei   \scriptscriptfont1\teni%
  \textfont2\seventeensy  \scriptfont2\twelvesy  \scriptscriptfont2\tensy%
  \textfont\itfam=\seventeenit\def\it{\fam\itfam\seventeenit}%
  \ifprod@font
    \scriptfont\itfam=\twelveit
      \scriptscriptfont\itfam=\tenit
  \else
    \scriptfont\itfam=\seventeenit
      \scriptscriptfont\itfam=\seventeenit
  \fi
  \textfont\bffam=\seventeenbf%
    \scriptfont\bffam=\twelvebf%
      \scriptscriptfont\bffam=\tenbf%
  \def\bf{\fam\bffam\seventeenbf}%
  \textfont\slfam=\seventeensl\def\sl{\fam\slfam\seventeensl}%
  \ifprod@font
    \scriptfont\slfam=\twelvesl
      \scriptscriptfont\slfam=\tensl
  \else
    \scriptfont\slfam=\seventeensl
      \scriptscriptfont\slfam=\seventeensl
  \fi
  \textfont\ttfam=\seventeentt\def\tt{\fam\ttfam\seventeentt}%
  \ifprod@font
    \scriptfont\ttfam=\twelvett
      \scriptscriptfont\ttfam=\tentt
  \else
    \scriptfont\ttfam=\seventeentt
      \scriptscriptfont\ttfam=\seventeentt
  \fi
  \textfont\scfam=\seventeencsc\def\sc{\fam\scfam\seventeencsc}%
  \ifprod@font
    \scriptfont\scfam=\twelvecsc
      \scriptscriptfont\scfam=\tencsc
  \else
    \scriptfont\scfam=\seventeencsc
      \scriptscriptfont\scfam=\seventeencsc
  \fi
  \textfont\sffam=\seventeensf\def\sf{\fam\sffam\seventeensf}%
  \ifprod@font
    \scriptfont\sffam=\twelvesf
      \scriptscriptfont\sffam=\tensf
  \else
    \scriptfont\sffam=\seventeensf
      \scriptscriptfont\sffam=\seventeensf
  \fi
  \textfont\mibfam=\seventeenmib
    \scriptfont\mibfam=\twelvemib
      \scriptscriptfont\mibfam=\tenmib
  \textfont\sybfam=\seventeensyb
    \scriptfont\sybfam=\twelvesyb
      \scriptscriptfont\sybfam=\tensyb
  \ifprod@font
    \textfont\xmfam=\seventeenxm
      \scriptfont\xmfam=\twelvexm
        \scriptscriptfont\xmfam=\tenxm
    \textfont\ymfam=\seventeenym
      \scriptfont\ymfam=\twelveym
        \scriptscriptfont\ymfam=\tenym
  \fi
  \def\oldstyle{\fam\@ne\seventeeni}%
  \def\boldstyle{\fam\mibfam\seventeenmib}%
  \b@ls{20pt}\rm%
}

\lineskip=1pt      \normallineskip=\lineskip
\lineskiplimit=\z@ \normallineskiplimit=\lineskiplimit



\def\la{\mathrel{\mathchoice {\vcenter{\offinterlineskip\halign{\hfil
$\displaystyle##$\hfil\cr<\cr\sim\cr}}}
{\vcenter{\offinterlineskip\halign{\hfil$\textstyle##$\hfil\cr
<\cr\sim\cr}}}
{\vcenter{\offinterlineskip\halign{\hfil$\scriptstyle##$\hfil\cr
<\cr\sim\cr}}}
{\vcenter{\offinterlineskip\halign{\hfil$\scriptscriptstyle##$\hfil\cr
<\cr\sim\cr}}}}}

\def\ga{\mathrel{\mathchoice {\vcenter{\offinterlineskip\halign{\hfil
$\displaystyle##$\hfil\cr>\cr\sim\cr}}}
{\vcenter{\offinterlineskip\halign{\hfil$\textstyle##$\hfil\cr
>\cr\sim\cr}}}
{\vcenter{\offinterlineskip\halign{\hfil$\scriptstyle##$\hfil\cr
>\cr\sim\cr}}}
{\vcenter{\offinterlineskip\halign{\hfil$\scriptscriptstyle##$\hfil\cr
>\cr\sim\cr}}}}}

\def\getsto{\mathrel{\mathchoice {\vcenter{\offinterlineskip
\halign{\hfil
$\displaystyle##$\hfil\cr\gets\cr\to\cr}}}
{\vcenter{\offinterlineskip\halign{\hfil$\textstyle##$\hfil\cr\gets
\cr\to\cr}}}
{\vcenter{\offinterlineskip\halign{\hfil$\scriptstyle##$\hfil\cr\gets
\cr\to\cr}}}
{\vcenter{\offinterlineskip\halign{\hfil$\scriptscriptstyle##$\hfil\cr
\gets\cr\to\cr}}}}}

\def\lid{\mathrel{\mathchoice {\vcenter{\offinterlineskip\halign{\hfil
$\displaystyle##$\hfil\cr<\cr\noalign{\vskip1.2pt}=\cr}}}
{\vcenter{\offinterlineskip\halign{\hfil$\textstyle##$\hfil\cr<\cr
\noalign{\vskip1.2pt}=\cr}}}
{\vcenter{\offinterlineskip\halign{\hfil$\scriptstyle##$\hfil\cr<\cr
\noalign{\vskip1pt}=\cr}}}
{\vcenter{\offinterlineskip\halign{\hfil$\scriptscriptstyle##$\hfil\cr
<\cr
\noalign{\vskip0.9pt}=\cr}}}}}

\def\gid{\mathrel{\mathchoice {\vcenter{\offinterlineskip\halign{\hfil
$\displaystyle##$\hfil\cr>\cr\noalign{\vskip1.2pt}=\cr}}}
{\vcenter{\offinterlineskip\halign{\hfil$\textstyle##$\hfil\cr>\cr
\noalign{\vskip1.2pt}=\cr}}}
{\vcenter{\offinterlineskip\halign{\hfil$\scriptstyle##$\hfil\cr>\cr
\noalign{\vskip1pt}=\cr}}}
{\vcenter{\offinterlineskip\halign{\hfil$\scriptscriptstyle##$\hfil\cr
>\cr
\noalign{\vskip0.9pt}=\cr}}}}}

\def\grole{\mathrel{\mathchoice {\vcenter{\offinterlineskip\halign{\hfil
$\displaystyle##$\hfil\cr>\cr\noalign{\vskip-1.5pt}<\cr}}}
{\vcenter{\offinterlineskip\halign{\hfil$\textstyle##$\hfil\cr
>\cr\noalign{\vskip-1.5pt}<\cr}}}
{\vcenter{\offinterlineskip\halign{\hfil$\scriptstyle##$\hfil\cr
>\cr\noalign{\vskip-1pt}<\cr}}}
{\vcenter{\offinterlineskip\halign{\hfil$\scriptscriptstyle##$\hfil\cr
>\cr\noalign{\vskip-0.5pt}<\cr}}}}}

\def\leogr{\mathrel{\mathchoice {\vcenter{\offinterlineskip\halign{\hfil
$\displaystyle##$\hfil\cr<\cr\noalign{\vskip-1.5pt}>\cr}}}
{\vcenter{\offinterlineskip\halign{\hfil$\textstyle##$\hfil\cr
<\cr\noalign{\vskip-1.5pt}>\cr}}}
{\vcenter{\offinterlineskip\halign{\hfil$\scriptstyle##$\hfil\cr
<\cr\noalign{\vskip-1pt}>\cr}}}
{\vcenter{\offinterlineskip\halign{\hfil$\scriptscriptstyle##$\hfil\cr
<\cr\noalign{\vskip-0.5pt}>\cr}}}}}

\def\loa{\mathrel{\mathchoice {\vcenter{\offinterlineskip\halign{\hfil
$\displaystyle##$\hfil\cr<\cr\approx\cr}}}
{\vcenter{\offinterlineskip\halign{\hfil$\textstyle##$\hfil\cr
<\cr\approx\cr}}}
{\vcenter{\offinterlineskip\halign{\hfil$\scriptstyle##$\hfil\cr
<\cr\approx\cr}}}
{\vcenter{\offinterlineskip\halign{\hfil$\scriptscriptstyle##$\hfil\cr
<\cr\approx\cr}}}}}

\def\goa{\mathrel{\mathchoice {\vcenter{\offinterlineskip\halign{\hfil
$\displaystyle##$\hfil\cr>\cr\approx\cr}}}
{\vcenter{\offinterlineskip\halign{\hfil$\textstyle##$\hfil\cr
>\cr\approx\cr}}}
{\vcenter{\offinterlineskip\halign{\hfil$\scriptstyle##$\hfil\cr
>\cr\approx\cr}}}
{\vcenter{\offinterlineskip\halign{\hfil$\scriptscriptstyle##$\hfil\cr
>\cr\approx\cr}}}}}

\def\diameter{{\ifmmode\mathchoice
{\ooalign{\hfil\hbox{$\displaystyle/$}\hfil\crcr
{\hbox{$\displaystyle\mathchar"20D$}}}}
{\ooalign{\hfil\hbox{$\textstyle/$}\hfil\crcr
{\hbox{$\textstyle\mathchar"20D$}}}}
{\ooalign{\hfil\hbox{$\scriptstyle/$}\hfil\crcr
{\hbox{$\scriptstyle\mathchar"20D$}}}}
{\ooalign{\hfil\hbox{$\scriptscriptstyle/$}\hfil\crcr
{\hbox{$\scriptscriptstyle\mathchar"20D$}}}}
\else{\ooalign{\hfil/\hfil\crcr\mathhexbox20D}}%
\fi}}

\def\sq{\ifmmode\squareforqed\else{\unskip\nobreak\hfil
\penalty50\hskip1em\null\nobreak\hfil\squareforqed
\parfillskip=0pt\finalhyphendemerits=0\endgraf}\fi}
\def\squareforqed{\hbox{\rlap{$\sqcap$}$\sqcup$}}


\def\bbbc{{\mathchoice {\setbox0=\hbox{$\displaystyle\rm C$}\hbox{\hbox
to0pt{\kern0.4\wd0\vrule height0.9\ht0\hss}\box0}}
{\setbox0=\hbox{$\textstyle\rm C$}\hbox{\hbox
to0pt{\kern0.4\wd0\vrule height0.9\ht0\hss}\box0}}
{\setbox0=\hbox{$\scriptstyle\rm C$}\hbox{\hbox
to0pt{\kern0.4\wd0\vrule height0.9\ht0\hss}\box0}}
{\setbox0=\hbox{$\scriptscriptstyle\rm C$}\hbox{\hbox
to0pt{\kern0.4\wd0\vrule height0.9\ht0\hss}\box0}}}}
\def\bbbq{{\mathchoice {\setbox0=\hbox{$\displaystyle\rm
Q$}\hbox{\raise
0.15\ht0\hbox to0pt{\kern0.4\wd0\vrule height0.8\ht0\hss}\box0}}
{\setbox0=\hbox{$\textstyle\rm Q$}\hbox{\raise
0.15\ht0\hbox to0pt{\kern0.4\wd0\vrule height0.8\ht0\hss}\box0}}
{\setbox0=\hbox{$\scriptstyle\rm Q$}\hbox{\raise
0.15\ht0\hbox to0pt{\kern0.4\wd0\vrule height0.7\ht0\hss}\box0}}
{\setbox0=\hbox{$\scriptscriptstyle\rm Q$}\hbox{\raise
0.15\ht0\hbox to0pt{\kern0.4\wd0\vrule height0.7\ht0\hss}\box0}}}}
\def\bbbt{{\mathchoice {\setbox0=\hbox{$\displaystyle\rm
T$}\hbox{\hbox to0pt{\kern0.3\wd0\vrule height0.9\ht0\hss}\box0}}
{\setbox0=\hbox{$\textstyle\rm T$}\hbox{\hbox
to0pt{\kern0.3\wd0\vrule height0.9\ht0\hss}\box0}}
{\setbox0=\hbox{$\scriptstyle\rm T$}\hbox{\hbox
to0pt{\kern0.3\wd0\vrule height0.9\ht0\hss}\box0}}
{\setbox0=\hbox{$\scriptscriptstyle\rm T$}\hbox{\hbox
to0pt{\kern0.3\wd0\vrule height0.9\ht0\hss}\box0}}}}
\def\bbbs{{\mathchoice
{\setbox0=\hbox{$\displaystyle     \rm S$}\hbox{\raise0.5\ht0\hbox
to0pt{\kern0.35\wd0\vrule height0.45\ht0\hss}\hbox
to0pt{\kern0.55\wd0\vrule height0.5\ht0\hss}\box0}}
{\setbox0=\hbox{$\textstyle        \rm S$}\hbox{\raise0.5\ht0\hbox
to0pt{\kern0.35\wd0\vrule height0.45\ht0\hss}\hbox
to0pt{\kern0.55\wd0\vrule height0.5\ht0\hss}\box0}}
{\setbox0=\hbox{$\scriptstyle      \rm S$}\hbox{\raise0.5\ht0\hbox
to0pt{\kern0.35\wd0\vrule height0.45\ht0\hss}\raise0.05\ht0\hbox
to0pt{\kern0.5\wd0\vrule height0.45\ht0\hss}\box0}}
{\setbox0=\hbox{$\scriptscriptstyle\rm S$}\hbox{\raise0.5\ht0\hbox
to0pt{\kern0.4\wd0\vrule height0.45\ht0\hss}\raise0.05\ht0\hbox
to0pt{\kern0.55\wd0\vrule height0.45\ht0\hss}\box0}}}}
\def\bbbz{{\mathchoice {\hbox{$\sf\textstyle Z\kern-0.4em Z$}}
{\hbox{$\sf\textstyle Z\kern-0.4em Z$}}
{\hbox{$\sf\scriptstyle Z\kern-0.3em Z$}}
{\hbox{$\sf\scriptscriptstyle Z\kern-0.2em Z$}}}}


\ifprod@font
  \mathchardef\la="3\@xm2E
  \mathchardef\getsto="3\@xm1C
  \mathchardef\lid="3\@xm35
  \mathchardef\grole="3\@xm3F
  \mathchardef\loa="3\@xm2F
  \mathchardef\ga="3\@xm26
  \mathchardef\gid="3\@xm3D
  \mathchardef\leogr="3\@xm37
  \mathchardef\goa="3\@xm27
  \mathchardef\sq="0\@xm03
%
%
\def\diameter{{%
  \ifmmode
    \mathchoice
    {\ooalign{\hfil\hbox{$\displaystyle/$}\hfil\crcr
    {\lower.2ex\hbox{$\displaystyle\mathchar"20D$}}}}%
    {\ooalign{\hfil\hbox{$\textstyle/$}\hfil\crcr
    {\lower.2ex\hbox{$\textstyle\mathchar"20D$}}}}%
    {\ooalign{\hfil\hbox{$\scriptstyle/$}\hfil\crcr
    {\lower.1ex\hbox{$\scriptstyle\mathchar"20D$}}}}%
    {\ooalign{\hfil\hbox{$\scriptscriptstyle/$}\hfil\crcr
    {\lower.1ex\hbox{$\scriptscriptstyle\mathchar"20D$}}}}%
  \else
    {\ooalign{\hfil/\hfil\crcr\lower.2ex\hbox{\mathhexbox20D}}}%
  \fi
}}
%
%

\def\bbbc{{\Bbb{C}}}
\def\bbbq{{\Bbb{Q}}}
\def\bbbt{{\Bbb{T}}}
\def\bbbs{{\Bbb{S}}}
\def\bbbz{{\Bbb{Z}}}
\fi


\ifprod@font
\mathchardef\boxdot="2\@xm00
\mathchardef\boxplus="2\@xm01
\mathchardef\boxtimes="2\@xm02
\mathchardef\square="0\@xm03
\mathchardef\blacksquare="0\@xm04
\mathchardef\centerdot="2\@xm05
\mathchardef\lozenge="0\@xm06
\mathchardef\blacklozenge="0\@xm07
\mathchardef\circlearrowright="3\@xm08
\mathchardef\circlearrowleft="3\@xm09
\mathchardef\rightleftharpoons="3\@xm0A
\mathchardef\leftrightharpoons="3\@xm0B
\mathchardef\boxminus="2\@xm0C
\mathchardef\Vdash="3\@xm0D
\mathchardef\Vvdash="3\@xm0E
\mathchardef\vDash="3\@xm0F
\mathchardef\twoheadrightarrow="3\@xm10
\mathchardef\twoheadleftarrow="3\@xm11
\mathchardef\leftleftarrows="3\@xm12
\mathchardef\rightrightarrows="3\@xm13
\mathchardef\upuparrows="3\@xm14
\mathchardef\downdownarrows="3\@xm15
\mathchardef\upharpoonright="3\@xm16

\mathchardef\downharpoonright="3\@xm17
\mathchardef\upharpoonleft="3\@xm18
\mathchardef\downharpoonleft="3\@xm19
\mathchardef\rightarrowtail="3\@xm1A
\mathchardef\leftarrowtail="3\@xm1B
\mathchardef\leftrightarrows="3\@xm1C
\mathchardef\rightleftarrows="3\@xm1D
\mathchardef\Lsh="3\@xm1E
\mathchardef\Rsh="3\@xm1F
\mathchardef\rightsquigarrow="3\@xm20
\mathchardef\leftrightsquigarrow="3\@xm21
\mathchardef\looparrowleft="3\@xm22
\mathchardef\looparrowright="3\@xm23
\mathchardef\circeq="3\@xm24
\mathchardef\succsim="3\@xm25
\mathchardef\gtrsim="3\@xm26
\mathchardef\gtrapprox="3\@xm27
\mathchardef\multimap="3\@xm28
\mathchardef\therefore="3\@xm29
\mathchardef\because="3\@xm2A
\mathchardef\doteqdot="3\@xm2B

\mathchardef\triangleq="3\@xm2C
\mathchardef\precsim="3\@xm2D
\mathchardef\lesssim="3\@xm2E
\mathchardef\lessapprox="3\@xm2F
\mathchardef\eqslantless="3\@xm30
\mathchardef\eqslantgtr="3\@xm31
\mathchardef\curlyeqprec="3\@xm32
\mathchardef\curlyeqsucc="3\@xm33
\mathchardef\preccurlyeq="3\@xm34
\mathchardef\leqq="3\@xm35
\mathchardef\leqslant="3\@xm36
\mathchardef\lessgtr="3\@xm37
\mathchardef\backprime="0\@xm38
\mathchardef\risingdotseq="3\@xm3A
\mathchardef\fallingdotseq="3\@xm3B
\mathchardef\succcurlyeq="3\@xm3C
\mathchardef\geqq="3\@xm3D
\mathchardef\geqslant="3\@xm3E
\mathchardef\gtrless="3\@xm3F
\mathchardef\sqsubset="3\@xm40
\mathchardef\sqsupset="3\@xm41
\mathchardef\vartriangleright="3\@xm42
\mathchardef\vartriangleleft="3\@xm43
\mathchardef\trianglerighteq="3\@xm44
\mathchardef\trianglelefteq="3\@xm45
\mathchardef\bigstar="0\@xm46
\mathchardef\between="3\@xm47
\mathchardef\blacktriangledown="0\@xm48
\mathchardef\blacktriangleright="3\@xm49
\mathchardef\blacktriangleleft="3\@xm4A
\mathchardef\vartriangle="0\@xm4D
\mathchardef\blacktriangle="0\@xm4E
\mathchardef\triangledown="0\@xm4F
\mathchardef\eqcirc="3\@xm50
\mathchardef\lesseqgtr="3\@xm51
\mathchardef\gtreqless="3\@xm52
\mathchardef\lesseqqgtr="3\@xm53
\mathchardef\gtreqqless="3\@xm54
\mathchardef\Rrightarrow="3\@xm56
\mathchardef\Lleftarrow="3\@xm57
\mathchardef\veebar="2\@xm59
\mathchardef\barwedge="2\@xm5A
\mathchardef\doublebarwedge="2\@xm5B
\mathchardef\angle="0\@xm5C
\mathchardef\measuredangle="0\@xm5D
\mathchardef\sphericalangle="0\@xm5E
\mathchardef\varpropto="3\@xm5F
\mathchardef\smallsmile="3\@xm60
\mathchardef\smallfrown="3\@xm61
\mathchardef\Subset="3\@xm62
\mathchardef\Supset="3\@xm63
\mathchardef\Cup="2\@xm64

\mathchardef\Cap="2\@xm65

\mathchardef\curlywedge="2\@xm66
\mathchardef\curlyvee="2\@xm67
\mathchardef\leftthreetimes="2\@xm68
\mathchardef\rightthreetimes="2\@xm69
\mathchardef\subseteqq="3\@xm6A
\mathchardef\supseteqq="3\@xm6B
\mathchardef\bumpeq="3\@xm6C
\mathchardef\Bumpeq="3\@xm6D
\mathchardef\lll="3\@xm6E

\mathchardef\ggg="3\@xm6F

\mathchardef\circledS="0\@xm73
\mathchardef\pitchfork="3\@xm74
\mathchardef\dotplus="2\@xm75
\mathchardef\backsim="3\@xm76
\mathchardef\backsimeq="3\@xm77
\mathchardef\complement="0\@xm7B
\mathchardef\intercal="2\@xm7C
\mathchardef\circledcirc="2\@xm7D
\mathchardef\circledast="2\@xm7E
\mathchardef\circleddash="2\@xm7F
\def\ulcorner{\delimiter"4\@xm70\@xm70 }
\def\urcorner{\delimiter"5\@xm71\@xm71 }
\def\llcorner{\delimiter"4\@xm78\@xm78 }
\def\lrcorner{\delimiter"5\@xm79\@xm79 }
\def\yen{\mathhexbox\@xm55 }
\def\checkmark{\mathhexbox\@xm58 }
\def\circledR{\mathhexbox\@xm72 }
\def\maltese{\mathhexbox\@xm7A }
\mathchardef\lvertneqq="3\@ym00
\mathchardef\gvertneqq="3\@ym01
\mathchardef\nleq="3\@ym02
\mathchardef\ngeq="3\@ym03
\mathchardef\nless="3\@ym04
\mathchardef\ngtr="3\@ym05
\mathchardef\nprec="3\@ym06
\mathchardef\nsucc="3\@ym07
\mathchardef\lneqq="3\@ym08
\mathchardef\gneqq="3\@ym09
\mathchardef\nleqslant="3\@ym0A
\mathchardef\ngeqslant="3\@ym0B
\mathchardef\lneq="3\@ym0C
\mathchardef\gneq="3\@ym0D
\mathchardef\npreceq="3\@ym0E
\mathchardef\nsucceq="3\@ym0F
\mathchardef\precnsim="3\@ym10
\mathchardef\succnsim="3\@ym11
\mathchardef\lnsim="3\@ym12
\mathchardef\gnsim="3\@ym13
\mathchardef\nleqq="3\@ym14
\mathchardef\ngeqq="3\@ym15
\mathchardef\precneqq="3\@ym16
\mathchardef\succneqq="3\@ym17
\mathchardef\precnapprox="3\@ym18
\mathchardef\succnapprox="3\@ym19
\mathchardef\lnapprox="3\@ym1A
\mathchardef\gnapprox="3\@ym1B
\mathchardef\nsim="3\@ym1C
\mathchardef\ncong="3\@ym1D

\mathchardef\varsubsetneq="3\@ym20
\mathchardef\varsupsetneq="3\@ym21
\mathchardef\nsubseteqq="3\@ym22
\mathchardef\nsupseteqq="3\@ym23
\mathchardef\subsetneqq="3\@ym24
\mathchardef\supsetneqq="3\@ym25
\mathchardef\varsubsetneqq="3\@ym26
\mathchardef\varsupsetneqq="3\@ym27
\mathchardef\subsetneq="3\@ym28
\mathchardef\supsetneq="3\@ym29
\mathchardef\nsubseteq="3\@ym2A
\mathchardef\nsupseteq="3\@ym2B
\mathchardef\nparallel="3\@ym2C
\mathchardef\nmid="3\@ym2D
\mathchardef\nshortmid="3\@ym2E
\mathchardef\nshortparallel="3\@ym2F
\mathchardef\nvdash="3\@ym30
\mathchardef\nVdash="3\@ym31
\mathchardef\nvDash="3\@ym32
\mathchardef\nVDash="3\@ym33
\mathchardef\ntrianglerighteq="3\@ym34
\mathchardef\ntrianglelefteq="3\@ym35
\mathchardef\ntriangleleft="3\@ym36
\mathchardef\ntriangleright="3\@ym37
\mathchardef\nleftarrow="3\@ym38
\mathchardef\nrightarrow="3\@ym39
\mathchardef\nLeftarrow="3\@ym3A
\mathchardef\nRightarrow="3\@ym3B
\mathchardef\nLeftrightarrow="3\@ym3C
\mathchardef\nleftrightarrow="3\@ym3D
\mathchardef\divideontimes="2\@ym3E
\mathchardef\varnothing="0\@ym3F
\mathchardef\nexists="0\@ym40
\mathchardef\mho="0\@ym66
\mathchardef\eth="0\@ym67
\mathchardef\eqsim="3\@ym68
\mathchardef\beth="0\@ym69
\mathchardef\gimel="0\@ym6A
\mathchardef\daleth="0\@ym6B
\mathchardef\lessdot="3\@ym6C
\mathchardef\gtrdot="3\@ym6D
\mathchardef\ltimes="2\@ym6E
\mathchardef\rtimes="2\@ym6F
\mathchardef\shortmid="3\@ym70
\mathchardef\shortparallel="3\@ym71
\mathchardef\smallsetminus="2\@ym72
\mathchardef\thicksim="3\@ym73
\mathchardef\thickapprox="3\@ym74
\mathchardef\approxeq="3\@ym75
\mathchardef\succapprox="3\@ym76
\mathchardef\precapprox="3\@ym77
\mathchardef\curvearrowleft="3\@ym78
\mathchardef\curvearrowright="3\@ym79
\mathchardef\digamma="0\@ym7A
\mathchardef\varkappa="0\@ym7B
\mathchardef\hslash="0\@ym7D
\mathchardef\hbar="0\@ym7E
\mathchardef\backepsilon="3\@ym7F


\def\Bbb{\ifmmode\let\next\Bbb@\else
\def\next{\errmessage{Use \string\Bbb\space only in math mode}}\fi\next}
\def\Bbb@#1{{\Bbb@@{#1}}}
\def\Bbb@@#1{\fam\ymfam#1}
\fi


\def\Nulle{0} 
\def\Afe{1}   
\def\Hae{2}   
\def\Hbe{3}   
\def\Hce{4}   
\def\Hde{5}   


\newcount\LastMac       \LastMac=\Nulle

\newskip\half      \half=5.5pt plus 1.5pt minus 2.25pt
\newskip\one       \one=11pt plus 3pt minus 5.5pt
\newskip\onehalf   \onehalf=16.5pt plus 5.5pt minus 8.25pt
\newskip\two       \two=22pt plus 5.5pt minus 11pt

\def\Half{\addvspace{\half}}
\def\One{\addvspace{\one}}
\def\OneHalf{\addvspace{\onehalf}}
\def\Two{\addvspace{\two}}


\def\Raggedright{
  \rightskip=\z@ plus \hsize\relax
}

\def\Fullout{
  \rightskip=\z@\relax
}

\def\Hang#1#2{
  \hangindent=#1%
  \hangafter=#2\relax
}


\newif\ifsp@page
\def\pagestyle#1{\csname ps@#1\endcsname}
\def\thispagestyle#1{\global\sp@pagetrue\gdef\sp@type{#1}}

\def\ps@titlepage{%
  \def\@oddhead{\eightpoint\noindent \the\CatchLine
    \ifprod@font\else\qquad Printed\ \today\fi \hfil}%
  \let\@evenhead=\@oddhead
}

\def\ps@headings{%
  \def\@oddhead{\elevenpoint\it\noindent
    \hfill\the\RightHeader\hskip1.5em\rm\folio}%
  \def\@evenhead{\elevenpoint\noindent
    \folio\hskip1.5em\it\the\LeftHeader\hfill}%
}

\def\ps@plate{%
  \def\@oddhead{\eightpoint\noindent\plt@cap\hfil}%
  \def\@evenhead{\eightpoint\noindent\plt@cap\hfil}%
}



\def\title#1{
  \bgroup
    \vbox to 8pt{\vss}%
    \seventeenpoint
    \Raggedright
    \noindent \strut{\bf #1}\par
  \egroup
}

\def\author#1{
  \bgroup
    \ifnum\LastMac=\Afe \OneHalf\else \vskip 21pt\fi
    \fourteenpoint
    \Raggedright
    \noindent \strut #1\par
    \vskip 3pt%
  \egroup
}

\def\affiliation#1{
  \bgroup
    \vskip -4pt%
    \eightpoint
    \Raggedright
    \noindent \strut {\it #1}\par
  \egroup
  \LastMac=\Afe\relax
}

\def\acceptedline#1{
  \bgroup
    \Two
    \eightpoint
    \Raggedright
    \noindent \strut #1\par
  \egroup
}

\long\def\abstract#1{%
  \bgroup
    \vskip 20pt%
    \everypar{\Hang{11pc}{0}}%
    \noindent{\ninebf ABSTRACT}\par
    \tenpoint
    \Fullout
    \noindent #1\par
  \egroup
}

\long\def\keywords#1{
  \bgroup
    \Half
    \everypar{\Hang{11pc}{0}}%
    \tenpoint
    \Fullout
    \noindent\hbox{\bf Key words:}\ #1\par
  \egroup
}


\def\maketitle{%
  \EndOpening
  \ifsinglecol \else \MakePage\fi
}


\def\pageoffset#1#2{\hoffset=#1\relax\voffset=#2\relax}


\def\Autonumber{
  \global\AutoNumbertrue  
}

\newif\ifAutoNumber \AutoNumberfalse
\newcount\Sec        
\newcount\SecSec
\newcount\SecSecSec

\Sec=\z@

\def\:{\let\@sptoken= } \:  
\def\:{\@xifnch} \expandafter\def\: {\futurelet\@tempc\@ifnch}

\def\@ifnextchar#1#2#3{%
  \let\@tempMACe #1%
  \def\@tempMACa{#2}%
  \def\@tempMACb{#3}%
  \futurelet \@tempMACc\@ifnch%
}

\def\@ifnch{%
\ifx \@tempMACc \@sptoken%
  \let\@tempMACd\@xifnch%
\else%
  \ifx \@tempMACc \@tempMACe%
    \let\@tempMACd\@tempMACa%
  \else%
    \let\@tempMACd\@tempMACb%
  \fi%
\fi%
\@tempMACd%
}

\def\@ifstar#1#2{\@ifnextchar *{\def\@tempMACa*{#1}\@tempMACa}{#2}}

\newskip\@tempskipb

\def\addvspace#1{%
  \ifvmode\else \endgraf\fi%
  \ifdim\lastskip=\z@%
    \vskip #1\relax%
  \else%
    \@tempskipb#1\relax\@xaddvskip%
  \fi%
}

\def\@xaddvskip{%
  \ifdim\lastskip<\@tempskipb%
    \vskip-\lastskip%
    \vskip\@tempskipb\relax%
  \else%
    \ifdim\@tempskipb<\z@%
      \ifdim\lastskip<\z@ \else%
        \advance\@tempskipb\lastskip%
        \vskip-\lastskip\vskip\@tempskipb%
      \fi%
    \fi%
  \fi%
}

\newskip\@tmpSKIP

\def\addpen#1{%
  \ifvmode
    \if@nobreak
    \else
      \ifdim\lastskip=\z@
        \penalty#1\relax
      \else
        \@tmpSKIP=\lastskip
        \vskip -\lastskip
        \penalty#1\vskip\@tmpSKIP
      \fi
    \fi
  \fi
}

\newcount\@clubpen   \@clubpen=\clubpenalty
\newif\if@nobreak    \@nobreakfalse

\def\@noafterindent{%
  \global\@nobreaktrue
  \everypar{\if@nobreak
              \global\@nobreakfalse
              \clubpenalty \@M
              {\setbox\z@\lastbox}%
              \LastMac=\Nulle\relax%
            \else
              \clubpenalty \@clubpen
              \everypar{}%
            \fi}
}

\newcount\gds@cbrk   \gds@cbrk=-300

\def\@nohdbrk{\interlinepenalty \@M\relax}

\let\@par=\par
\def\@restorepar{\def\par{\@par}}

\newif\if@endpe   \@endpefalse
 
\def\@doendpe{\@endpetrue \@nobreakfalse \LastMac=\Nulle\relax%
     \def\par{\@restorepar\everypar{}\par\@endpefalse}%
              \everypar{\setbox\z@\lastbox\everypar{}\@endpefalse}%
}

\def\section{\@ifstar{\@ssection}{\@section}}

\def\@section#1{
  \if@nobreak
    \everypar{}%
    \ifnum\LastMac=\Hae \addvspace{\half}\fi
  \else
    \addpen{\gds@cbrk}%
    \addvspace{\two}%
  \fi
  \bgroup
    \ninepoint\bf
    \Raggedright
    \ifAutoNumber
      \global\advance\Sec \@ne
      \noindent\@nohdbrk\number\Sec\hskip 1pc \uppercase{#1}\par
      \global\SecSec=\z@
    \else
      \noindent\@nohdbrk\uppercase{#1}\par
    \fi
  \egroup
  \nobreak
  \vskip\half
  \nobreak
  \@noafterindent
  \LastMac=\Hae\relax
}

\def\@ssection#1{
  \if@nobreak
    \everypar{}%
    \ifnum\LastMac=\Hae \addvspace{\half}\fi
  \else
    \addpen{\gds@cbrk}%
    \addvspace{\two}%
  \fi
  \bgroup
    \ninepoint\bf
    \Raggedright
    \noindent\@nohdbrk\uppercase{#1}\par
  \egroup
  \nobreak
  \vskip\half
  \nobreak
  \@noafterindent
  \LastMac=\Hae\relax
}

\def\subsection#1{
  \if@nobreak
    \everypar{}%
    \ifnum\LastMac=\Hae \addvspace{1pt plus 1pt minus .5pt}\fi
  \else
    \addpen{\gds@cbrk}%
    \addvspace{\onehalf}%
  \fi
  \bgroup
    \ninepoint\bf
    \Raggedright
    \ifAutoNumber
      \global\advance\SecSec \@ne
      \noindent\@nohdbrk\number\Sec.\number\SecSec \hskip 1pc\relax #1\par
      \global\SecSecSec=\z@
    \else
      \noindent\@nohdbrk #1\par
    \fi
  \egroup
  \nobreak
  \vskip\half
  \nobreak
  \@noafterindent
  \LastMac=\Hbe\relax
}

\def\subsubsection#1{
  \if@nobreak
    \everypar{}%
    \ifnum\LastMac=\Hbe \addvspace{1pt plus 1pt minus .5pt}\fi
  \else
    \addpen{\gds@cbrk}%
    \addvspace{\onehalf}%
  \fi
  \bgroup
    \ninepoint\it
    \Raggedright
    \ifAutoNumber
      \global\advance\SecSecSec \@ne
      \noindent\@nohdbrk\number\Sec.\number\SecSec.\number\SecSecSec
        \hskip 1pc\relax #1\par
    \else
      \noindent\@nohdbrk #1\par
    \fi
  \egroup
  \nobreak
  \vskip\half
  \nobreak
  \@noafterindent
  \LastMac=\Hce\relax
}

\def\paragraph#1{
  \if@nobreak
    \everypar{}%
  \else
    \addpen{\gds@cbrk}%
    \addvspace{\one}%
  \fi%
  \bgroup%
    \ninepoint\it
    \noindent #1\ \nobreak%
  \egroup
  \LastMac=\Hde\relax
  \ignorespaces
}




\def\beginlist{%
  \par\if@nobreak \else\addvspace{\half}\fi%
  \bgroup%
    \ninepoint
    \let\item=\list@item%
}

\def\list@item{%
  \par\noindent\hskip 1em\relax%
  \ignorespaces%
}

\def\endlist{\par\egroup\addvspace{\half}\@doendpe}


\def\beginrefs{%
  \par
  \bgroup
    \eightpoint
    \Raggedright
    \let\bibitem=\bib@item
}

\def\bib@item{%
  \par\parindent=1.5em\Hang{1.5em}{1}%
  \everypar={\Hang{1.5em}{1}\ignorespaces}%
  \noindent\ignorespaces
}

\def\endrefs{\par\egroup\@doendpe}


\newtoks\CatchLine

\def\@journal{Mon.\ Not.\ R.\ Astron.\ Soc.\ }  
\def\@pubyear{1994}        
\def\@pagerange{000--000}  
\def\@volume{000}          
\def\@microfiche{}         %

\def\pubyear#1{\gdef\@pubyear{#1}\@makecatchline}
\def\pagerange#1{\gdef\@pagerange{#1}\@makecatchline}
\def\volume#1{\gdef\@volume{#1}\@makecatchline}
\def\microfiche#1{\gdef\@microfiche{and Microfiche\ #1}\@makecatchline}

\def\@makecatchline{%
  \global\CatchLine{%
    {\rm \@journal {\bf \@volume},\ \@pagerange\ (\@pubyear)\ \@microfiche}}%
}

\@makecatchline 

\newtoks\LeftHeader
\def\shortauthor#1{
  \global\LeftHeader{#1}%
}

\newtoks\RightHeader
\def\shorttitle#1{
  \global\RightHeader{#1}%
}

\def\PageHead{
  \begingroup
    \ifsp@page
      \csname ps@\sp@type\endcsname
      \global\sp@pagefalse
    \fi
    \ifodd\pageno
      \let\the@head=\@oddhead
    \else
      \let\the@head=\@evenhead
    \fi
    \vbox to \z@{\vskip-22.5\p@%
      \hbox to \PageWidth{\vbox to8.5\p@{}%
        \the@head
      }%
    \vss}%
  \endgroup
  \nointerlineskip
}

\def\today{%
  \number\day\space
  \ifcase\month\or January\or February\or March\or April\or May\or June\or
    July\or August\or September\or October\or November\or December\fi
  \space\number\year%
}

\def\PageFoot{} 

\def\authorcomment#1{%
  \gdef\PageFoot{%
    \nointerlineskip%
    \vbox to 22pt{\vfil%
      \hbox to \PageWidth{\elevenpoint\noindent \hfil #1 \hfil}}%
  }%
}


\newif\ifplate@page
\newbox\plt@box

\def\beginplatepage{%
  \let\plate=\plate@head
  \let\caption=\fig@caption
  \global\setbox\plt@box=\vbox\bgroup
  \TEMPDIMEN=\PageWidth 
  \hsize=\PageWidth\relax
}

\def\endplatepage{\par\egroup\global\plate@pagetrue}
\def\plate@head#1{\gdef\plt@cap{#1}}


\def\letters{%
  \gdef\folio{\ifnum\pageno<\z@ L\romannumeral-\pageno
    \else L\number\pageno \fi}%
}


\everydisplay{\displaysetup}

\newif\ifeqno
\newif\ifleqno

\def\displaysetup#1$${%
 \displaytest#1\eqno\eqno\displaytest
}

\def\displaytest#1\eqno#2\eqno#3\displaytest{%
 \if!#3!\ldisplaytest#1\leqno\leqno\ldisplaytest
 \else\eqnotrue\leqnofalse\def\eqn{#2}\def\eq{#1}\fi
 \generaldisplay$$}

\def\ldisplaytest#1\leqno#2\leqno#3\ldisplaytest{%
 \def\eq{#1}%
 \if!#3!\eqnofalse\else\eqnotrue\leqnotrue
  \def\eqn{#2}\fi}

\def\generaldisplay{%
\ifeqno \ifleqno 
   \hbox to \hsize{\noindent
     $\displaystyle\eq$\hfil$\displaystyle\eqn$}
  \else
    \hbox to \hsize{\noindent
     $\displaystyle\eq$\hfil$\displaystyle\eqn$}
  \fi
 \else
 \hbox to \hsize{\vbox{\noindent
  $\displaystyle\eq$\hfil}}
 \fi
}


\def\@notice{%
  \par\Two%
  \noindent{\b@ls{11pt}\ninerm This paper has been produced using the
    Blackwell Scientific Publications \TeX\ macros.\par}%
}

\outer\def\bye{\@notice\par\vfill\supereject\end}


\def\start@mess{%
  Monthly notices of the RAS journal style (\@typeface)\space
    v\@version,\space \@verdate.%
}

\everyjob{\Warn{\start@mess}}



\newif\if@debug \@debugfalse  

\def\Print#1{\if@debug\immediate\write16{#1}\else \fi}
\def\Warn#1{\immediate\write16{#1}}
\def\wlog#1{}

\newcount\Iteration 

\def\Single{0} \def\Double{1}                 
\def\Figure{0} \def\Table{1}                  

\def\InStack{0}  
\def\InZoneA{1}
\def\InZoneB{2}
\def\InZoneC{3}

\newcount\TEMPCOUNT 
\newdimen\TEMPDIMEN 
\newbox\TEMPBOX     
\newbox\VOIDBOX     

\newcount\LengthOfStack 
\newcount\MaxItems      
\newcount\StackPointer
\newcount\Point         
\newcount\NextFigure    
\newcount\NextTable     
\newcount\NextItem      

\newcount\StatusStack   
\newcount\NumStack      
\newcount\TypeStack     
\newcount\SpanStack     
\newcount\BoxStack      

\newcount\ItemSTATUS    
\newcount\ItemNUMBER    
\newcount\ItemTYPE      
\newcount\ItemSPAN      
\newbox\ItemBOX         
\newdimen\ItemSIZE      

\newdimen\PageHeight    
\newdimen\TextLeading   
\newdimen\Feathering    
\newcount\LinesPerPage  
\newdimen\ColumnWidth   
\newdimen\ColumnGap     
\newdimen\PageWidth     
\newdimen\BodgeHeight   
\newcount\Leading       

\newdimen\ZoneBSize  
\newdimen\TextSize   
\newbox\ZoneABOX     
\newbox\ZoneBBOX     
\newbox\ZoneCBOX     

\newif\ifFirstSingleItem
\newif\ifFirstZoneA
\newif\ifMakePageInComplete
\newif\ifMoreFigures \MoreFiguresfalse 
\newif\ifMoreTables  \MoreTablesfalse  

\newif\ifFigInZoneB 
\newif\ifFigInZoneC 
\newif\ifTabInZoneB 
\newif\ifTabInZoneC

\newif\ifZoneAFullPage

\newbox\MidBOX    
\newbox\LeftBOX
\newbox\RightBOX
\newbox\PageBOX   

\newif\ifLeftCOL  
\LeftCOLtrue

\newdimen\ZoneBAdjust

\newcount\ItemFits
\def\Yes{1}
\def\No{2}


\MaxItems=15
\NextFigure=\z@        
\NextTable=\@ne

\BodgeHeight=6pt
\TextLeading=11pt    
\Leading=11
\Feathering=\z@      
\LinesPerPage=61     
\topskip=\TextLeading
\ColumnWidth=20pc    
\ColumnGap=2pc       

\newskip\ItemSepamount  
\ItemSepamount=\TextLeading plus \TextLeading minus 4pt

\parskip=\z@ plus .1pt
\parindent=18pt
\widowpenalty=\z@
\clubpenalty=10000
\tolerance=1500
\hbadness=1500
\abovedisplayskip=6pt plus 2pt minus 2pt
\belowdisplayskip=6pt plus 2pt minus 2pt
\abovedisplayshortskip=6pt plus 2pt minus 2pt
\belowdisplayshortskip=6pt plus 2pt minus 2pt

\ninepoint 


\PageHeight=682pt

\PageWidth=2\ColumnWidth
\advance\PageWidth by \ColumnGap

\pagestyle{headings}




\newcount\DUMMY \StatusStack=\allocationnumber
\newcount\DUMMY \newcount\DUMMY \newcount\DUMMY 
\newcount\DUMMY \newcount\DUMMY \newcount\DUMMY 
\newcount\DUMMY \newcount\DUMMY \newcount\DUMMY
\newcount\DUMMY \newcount\DUMMY \newcount\DUMMY 
\newcount\DUMMY \newcount\DUMMY \newcount\DUMMY

\newcount\DUMMY \NumStack=\allocationnumber
\newcount\DUMMY \newcount\DUMMY \newcount\DUMMY 
\newcount\DUMMY \newcount\DUMMY \newcount\DUMMY 
\newcount\DUMMY \newcount\DUMMY \newcount\DUMMY 
\newcount\DUMMY \newcount\DUMMY \newcount\DUMMY 
\newcount\DUMMY \newcount\DUMMY \newcount\DUMMY

\newcount\DUMMY \TypeStack=\allocationnumber
\newcount\DUMMY \newcount\DUMMY \newcount\DUMMY 
\newcount\DUMMY \newcount\DUMMY \newcount\DUMMY 
\newcount\DUMMY \newcount\DUMMY \newcount\DUMMY 
\newcount\DUMMY \newcount\DUMMY \newcount\DUMMY 
\newcount\DUMMY \newcount\DUMMY \newcount\DUMMY

\newcount\DUMMY \SpanStack=\allocationnumber
\newcount\DUMMY \newcount\DUMMY \newcount\DUMMY 
\newcount\DUMMY \newcount\DUMMY \newcount\DUMMY 
\newcount\DUMMY \newcount\DUMMY \newcount\DUMMY 
\newcount\DUMMY \newcount\DUMMY \newcount\DUMMY 
\newcount\DUMMY \newcount\DUMMY \newcount\DUMMY

\newbox\DUMMY   \BoxStack=\allocationnumber
\newbox\DUMMY   \newbox\DUMMY \newbox\DUMMY 
\newbox\DUMMY   \newbox\DUMMY \newbox\DUMMY 
\newbox\DUMMY   \newbox\DUMMY \newbox\DUMMY 
\newbox\DUMMY   \newbox\DUMMY \newbox\DUMMY 
\newbox\DUMMY   \newbox\DUMMY \newbox\DUMMY

\def\wlog{\immediate\write\m@ne}


\def\GetItemAll#1{%
 \GetItemSTATUS{#1}
 \GetItemNUMBER{#1}
 \GetItemTYPE{#1}
 \GetItemSPAN{#1}
 \GetItemBOX{#1}
}

\def\GetItemSTATUS#1{%
 \Point=\StatusStack
 \advance\Point by #1
 \global\ItemSTATUS=\count\Point
}

\def\GetItemNUMBER#1{%
 \Point=\NumStack
 \advance\Point by #1
 \global\ItemNUMBER=\count\Point
}

\def\GetItemTYPE#1{%
 \Point=\TypeStack
 \advance\Point by #1
 \global\ItemTYPE=\count\Point
}

\def\GetItemSPAN#1{%
 \Point\SpanStack
 \advance\Point by #1
 \global\ItemSPAN=\count\Point
}

\def\GetItemBOX#1{%
 \Point=\BoxStack
 \advance\Point by #1
 \global\setbox\ItemBOX=\vbox{\copy\Point}
 \global\ItemSIZE=\ht\ItemBOX
 \global\advance\ItemSIZE by \dp\ItemBOX
 \TEMPCOUNT=\ItemSIZE
 \divide\TEMPCOUNT by \Leading
 \divide\TEMPCOUNT by 65536
 \advance\TEMPCOUNT \@ne
 \ItemSIZE=\TEMPCOUNT pt
 \global\multiply\ItemSIZE by \Leading
}


\def\JoinStack{%
 \ifnum\LengthOfStack=\MaxItems 
  \Warn{WARNING: Stack is full...some items will be lost!}
 \else
  \Point=\StatusStack
  \advance\Point by \LengthOfStack
  \global\count\Point=\ItemSTATUS
  \Point=\NumStack
  \advance\Point by \LengthOfStack
  \global\count\Point=\ItemNUMBER
  \Point=\TypeStack
  \advance\Point by \LengthOfStack
  \global\count\Point=\ItemTYPE
  \Point\SpanStack
  \advance\Point by \LengthOfStack
  \global\count\Point=\ItemSPAN
  \Point=\BoxStack
  \advance\Point by \LengthOfStack
  \global\setbox\Point=\vbox{\copy\ItemBOX}
  \global\advance\LengthOfStack \@ne
  \ifnum\ItemTYPE=\Figure 
   \global\MoreFigurestrue
  \else
   \global\MoreTablestrue
  \fi
 \fi
}


\def\LeaveStack#1{%
 {\Iteration=#1
 \loop
 \ifnum\Iteration<\LengthOfStack
  \advance\Iteration \@ne
  \GetItemSTATUS{\Iteration}
   \advance\Point by \m@ne
   \global\count\Point=\ItemSTATUS
  \GetItemNUMBER{\Iteration}
   \advance\Point by \m@ne
   \global\count\Point=\ItemNUMBER
  \GetItemTYPE{\Iteration}
   \advance\Point by \m@ne
   \global\count\Point=\ItemTYPE
  \GetItemSPAN{\Iteration}
   \advance\Point by \m@ne
   \global\count\Point=\ItemSPAN
  \GetItemBOX{\Iteration}
   \advance\Point by \m@ne
   \global\setbox\Point=\vbox{\copy\ItemBOX}
 \repeat}
 \global\advance\LengthOfStack by \m@ne
}


\newif\ifStackNotClean

\def\CleanStack{%
 \StackNotCleantrue
 {\Iteration=\z@
  \loop
   \ifStackNotClean
    \GetItemSTATUS{\Iteration}
    \ifnum\ItemSTATUS=\InStack
     \advance\Iteration \@ne
     \else
      \LeaveStack{\Iteration}
    \fi
   \ifnum\LengthOfStack<\Iteration
    \StackNotCleanfalse
   \fi
 \repeat}
}


\def\FindItem#1#2{%
 \global\StackPointer=\m@ne 
 {\Iteration=\z@
  \loop
  \ifnum\Iteration<\LengthOfStack
   \GetItemSTATUS{\Iteration}
   \ifnum\ItemSTATUS=\InStack
    \GetItemTYPE{\Iteration}
    \ifnum\ItemTYPE=#1
     \GetItemNUMBER{\Iteration}
     \ifnum\ItemNUMBER=#2
      \global\StackPointer=\Iteration
      \Iteration=\LengthOfStack 
     \fi
    \fi
   \fi
  \advance\Iteration \@ne
 \repeat}
}


\def\FindNext{%
 \global\StackPointer=\m@ne 
 {\Iteration=\z@
  \loop
  \ifnum\Iteration<\LengthOfStack
   \GetItemSTATUS{\Iteration}
   \ifnum\ItemSTATUS=\InStack
    \GetItemTYPE{\Iteration}
   \ifnum\ItemTYPE=\Figure
    \ifMoreFigures
      \global\NextItem=\Figure
      \global\StackPointer=\Iteration
      \Iteration=\LengthOfStack 
    \fi
   \fi
   \ifnum\ItemTYPE=\Table
    \ifMoreTables
      \global\NextItem=\Table
      \global\StackPointer=\Iteration
      \Iteration=\LengthOfStack 
    \fi
   \fi
  \fi
  \advance\Iteration \@ne
 \repeat}
}


\def\ChangeStatus#1#2{%
 \Point=\StatusStack
 \advance\Point by #1
 \global\count\Point=#2
}



\def\Zone{\InZoneA}

\ZoneBAdjust=\z@

\def\MakePage{
 \global\ZoneBSize=\PageHeight
 \global\TextSize=\ZoneBSize
 \global\ZoneAFullPagefalse
 \global\topskip=\TextLeading
 \MakePageInCompletetrue
 \MoreFigurestrue
 \MoreTablestrue
 \FigInZoneBfalse
 \FigInZoneCfalse
 \TabInZoneBfalse
 \TabInZoneCfalse
 \global\FirstSingleItemtrue
 \global\FirstZoneAtrue
 \global\setbox\ZoneABOX=\box\VOIDBOX
 \global\setbox\ZoneBBOX=\box\VOIDBOX
 \global\setbox\ZoneCBOX=\box\VOIDBOX
 \loop
  \ifMakePageInComplete
 \FindNext
 \ifnum\StackPointer=\m@ne
  \NextItem=\m@ne
  \MoreFiguresfalse
  \MoreTablesfalse
 \fi
 \ifnum\NextItem=\Figure
   \FindItem{\Figure}{\NextFigure}
   \ifnum\StackPointer=\m@ne \global\MoreFiguresfalse
   \else
    \GetItemSPAN{\StackPointer}
    \ifnum\ItemSPAN=\Single \def\Zone{\InZoneB}\relax
     \ifFigInZoneC \global\MoreFiguresfalse\fi
    \else
     \def\Zone{\InZoneA}
     \ifFigInZoneB \def\Zone{\InZoneC}\fi
    \fi
   \fi
   \ifMoreFigures\Print{}\FigureItems\fi
 \fi
\ifnum\NextItem=\Table
   \FindItem{\Table}{\NextTable}
   \ifnum\StackPointer=\m@ne \global\MoreTablesfalse
   \else
    \GetItemSPAN{\StackPointer}
    \ifnum\ItemSPAN=\Single\relax
     \ifTabInZoneC \global\MoreTablesfalse\fi
    \else
     \def\Zone{\InZoneA}
     \ifTabInZoneB \def\Zone{\InZoneC}\fi
    \fi
   \fi
   \ifMoreTables\Print{}\TableItems\fi
 \fi
   \MakePageInCompletefalse 
   \ifMoreFigures\MakePageInCompletetrue\fi
   \ifMoreTables\MakePageInCompletetrue\fi
 \repeat
 \ifZoneAFullPage
  \global\TextSize=\z@
  \global\ZoneBSize=\z@
  \global\vsize=\z@\relax
  \global\topskip=\z@\relax
  \vbox to \z@{\vss}
  \eject
 \else
 \global\advance\ZoneBSize by -\ZoneBAdjust
 \global\vsize=\ZoneBSize
 \global\hsize=\ColumnWidth
 \global\ZoneBAdjust=\z@
 \ifdim\TextSize<23pt
 \Warn{}
 \Warn{* Making column fall short: TextSize=\the\TextSize *}
 \vskip-\lastskip\eject\fi
 \fi
}

\def\MakeRightCol{
 \global\TextSize=\ZoneBSize
 \MakePageInCompletetrue
 \MoreFigurestrue
 \MoreTablestrue
 \global\FirstSingleItemtrue
 \global\setbox\ZoneBBOX=\box\VOIDBOX
 \def\Zone{\InZoneB}
 \loop
  \ifMakePageInComplete
 \FindNext
 \ifnum\StackPointer=\m@ne
  \NextItem=\m@ne
  \MoreFiguresfalse
  \MoreTablesfalse
 \fi
 \ifnum\NextItem=\Figure
   \FindItem{\Figure}{\NextFigure}
   \ifnum\StackPointer=\m@ne \MoreFiguresfalse
   \else
    \GetItemSPAN{\StackPointer}
    \ifnum\ItemSPAN=\Double\relax
     \MoreFiguresfalse\fi
   \fi
   \ifMoreFigures\Print{}\FigureItems\fi
 \fi
 \ifnum\NextItem=\Table
   \FindItem{\Table}{\NextTable}
   \ifnum\StackPointer=\m@ne \MoreTablesfalse
   \else
    \GetItemSPAN{\StackPointer}
    \ifnum\ItemSPAN=\Double\relax
     \MoreTablesfalse\fi
   \fi
   \ifMoreTables\Print{}\TableItems\fi
 \fi
   \MakePageInCompletefalse 
   \ifMoreFigures\MakePageInCompletetrue\fi
   \ifMoreTables\MakePageInCompletetrue\fi
 \repeat
 \ifZoneAFullPage
  \global\TextSize=\z@
  \global\ZoneBSize=\z@
  \global\vsize=\z@\relax
  \global\topskip=\z@\relax
  \vbox to \z@{\vss}
  \eject
 \else
 \global\vsize=\ZoneBSize
 \global\hsize=\ColumnWidth
 \ifdim\TextSize<23pt
 \Warn{}
 \Warn{* Making column fall short: TextSize=\the\TextSize *}
 \vskip-\lastskip\eject\fi
\fi
}

\def\FigureItems{
 \Print{Considering...}
 \ShowItem{\StackPointer}
 \GetItemBOX{\StackPointer} 
 \GetItemSPAN{\StackPointer}
  \CheckFitInZone 
  \ifnum\ItemFits=\Yes
   \ifnum\ItemSPAN=\Single
     \ChangeStatus{\StackPointer}{\InZoneB} 
     \global\FigInZoneBtrue
     \ifFirstSingleItem
      \hbox{}\vskip-\BodgeHeight
     \global\advance\ItemSIZE by \TextLeading
     \fi
     \unvbox\ItemBOX\ItemSep
     \global\FirstSingleItemfalse
     \global\advance\TextSize by -\ItemSIZE
     \global\advance\TextSize by -\TextLeading
   \else
    \ifFirstZoneA
     \global\advance\ItemSIZE by \TextLeading
     \global\FirstZoneAfalse\fi
    \global\advance\TextSize by -\ItemSIZE
    \global\advance\TextSize by -\TextLeading
    \global\advance\ZoneBSize by -\ItemSIZE
    \global\advance\ZoneBSize by -\TextLeading
    \ifFigInZoneB\relax
     \else
     \ifdim\TextSize<3\TextLeading
     \global\ZoneAFullPagetrue
     \fi
    \fi
    \ChangeStatus{\StackPointer}{\Zone}
    \ifnum\Zone=\InZoneC \global\FigInZoneCtrue\fi
  \fi
   \Print{TextSize=\the\TextSize}
   \Print{ZoneBSize=\the\ZoneBSize}
  \global\advance\NextFigure \@ne
   \Print{This figure has been placed.}
  \else
   \Print{No space available for this figure...holding over.}
   \Print{}
   \global\MoreFiguresfalse
  \fi
}

\def\TableItems{
 \Print{Considering...}
 \ShowItem{\StackPointer}
 \GetItemBOX{\StackPointer} 
 \GetItemSPAN{\StackPointer}
  \CheckFitInZone 
  \ifnum\ItemFits=\Yes
   \ifnum\ItemSPAN=\Single
    \ChangeStatus{\StackPointer}{\InZoneB}
     \global\TabInZoneBtrue
     \ifFirstSingleItem
      \hbox{}\vskip-\BodgeHeight
     \global\advance\ItemSIZE by \TextLeading
     \fi
     \unvbox\ItemBOX\ItemSep
     \global\FirstSingleItemfalse
     \global\advance\TextSize by -\ItemSIZE
     \global\advance\TextSize by -\TextLeading
   \else
    \ifFirstZoneA
    \global\advance\ItemSIZE by \TextLeading
    \global\FirstZoneAfalse\fi
    \global\advance\TextSize by -\ItemSIZE
    \global\advance\TextSize by -\TextLeading
    \global\advance\ZoneBSize by -\ItemSIZE
    \global\advance\ZoneBSize by -\TextLeading
    \ifFigInZoneB\relax
     \else
     \ifdim\TextSize<3\TextLeading
     \global\ZoneAFullPagetrue
     \fi
    \fi
    \ChangeStatus{\StackPointer}{\Zone}
    \ifnum\Zone=\InZoneC \global\TabInZoneCtrue\fi
   \fi
  \global\advance\NextTable \@ne
   \Print{This table has been placed.}
  \else
  \Print{No space available for this table...holding over.}
   \Print{}
   \global\MoreTablesfalse
  \fi
}


\def\CheckFitInZone{%
{\advance\TextSize by -\ItemSIZE
 \advance\TextSize by -\TextLeading
 \ifFirstSingleItem
  \advance\TextSize by \TextLeading
 \fi
 \ifnum\Zone=\InZoneA\relax
  \else \advance\TextSize by -\ZoneBAdjust
 \fi
 \ifdim\TextSize<3\TextLeading \global\ItemFits=\No
 \else \global\ItemFits=\Yes\fi}
}

\def\BeginOpening{%
  \thispagestyle{titlepage}%
  \global\setbox\ItemBOX=\vbox\bgroup%
    \hsize=\PageWidth%
    \hrule height \z@
    \ifsinglecol\vskip 6pt\fi 
}

\let\begintopmatter=\BeginOpening  

\def\EndOpening{%
  \One
  \egroup
  \ifsinglecol
    \box\ItemBOX%
    \vskip\TextLeading plus 2\TextLeading
    \@noafterindent
  \else
    \ItemNUMBER=\z@%
    \ItemTYPE=\Figure
    \ItemSPAN=\Double
    \ItemSTATUS=\InStack
    \JoinStack
  \fi
}


\newif\if@here  \@herefalse

\def\no@float{\global\@heretrue}
\let\nofloat=\relax 

\def\beginfigure{%
  \@ifstar{\global\@dfloattrue \@bfigure}{\global\@dfloatfalse \@bfigure}%
}

\def\@bfigure#1{%
  \par
  \if@dfloat
    \ItemSPAN=\Double
    \TEMPDIMEN=\PageWidth
  \else
    \ItemSPAN=\Single
    \TEMPDIMEN=\ColumnWidth
  \fi
  \ifsinglecol
    \TEMPDIMEN=\PageWidth
  \else
    \ItemSTATUS=\InStack
    \ItemNUMBER=#1%
    \ItemTYPE=\Figure
  \fi
  \bgroup
    \hsize=\TEMPDIMEN
    \global\setbox\ItemBOX=\vbox\bgroup
      \eightpoint\nostb@ls{10pt}%
      \let\caption=\fig@caption
      \ifsinglecol \let\nofloat=\no@float\fi
}

\def\fig@caption#1{%
  \vskip 5.5pt plus 6pt%
  \bgroup 
    \eightpoint\nostb@ls{10pt}%
    \setbox\TEMPBOX=\hbox{#1}%
    \ifdim\wd\TEMPBOX>\TEMPDIMEN
      \noindent \unhbox\TEMPBOX\par
    \else
      \hbox to \hsize{\hfil\unhbox\TEMPBOX\hfil}%
    \fi
  \egroup
}

\def\endfigure{%
  \par\egroup 
  \egroup
  \ifsinglecol
    \if@here \midinsert\global\@herefalse\else \topinsert\fi
      \unvbox\ItemBOX
    \endinsert
  \else
    \JoinStack
    \Print{Processing source for figure \the\ItemNUMBER}%
  \fi
}


\newbox\tab@cap@box
\def\tab@caption#1{\global\setbox\tab@cap@box=\hbox{#1\par}}

\newtoks\tab@txt@toks
\long\def\tab@txt#1{\global\tab@txt@toks={#1}\global\table@txttrue}

\newif\iftable@txt  \table@txtfalse
\newif\if@dfloat    \@dfloatfalse

\def\begintable{%
  \@ifstar{\global\@dfloattrue \@btable}{\global\@dfloatfalse \@btable}%
}

\def\@btable#1{%
  \par
  \if@dfloat
    \ItemSPAN=\Double
    \TEMPDIMEN=\PageWidth
  \else
    \ItemSPAN=\Single
    \TEMPDIMEN=\ColumnWidth
  \fi
  \ifsinglecol
    \TEMPDIMEN=\PageWidth
  \else
    \ItemSTATUS=\InStack
    \ItemNUMBER=#1%
    \ItemTYPE=\Table
  \fi
  \bgroup
    \eightpoint\nostb@ls{10pt}%
    \global\setbox\ItemBOX=\vbox\bgroup
      \let\caption=\tab@caption
      \let\tabletext=\tab@txt
      \ifsinglecol \let\nofloat=\no@float\fi
}

\def\endtable{%
  \par\egroup 
  \egroup
  \setbox\TEMPBOX=\hbox to \TEMPDIMEN{%
    \hss
    \vbox{%
      \hsize=\wd\ItemBOX
      \ifvoid\tab@cap@box
      \else
        \noindent\unhbox\tab@cap@box
        \vskip 5.5pt plus 6pt%
      \fi
      \box\ItemBOX
      \iftable@txt
        \vskip 10pt%
        \eightpoint\nostb@ls{10pt}%
        \noindent\the\tab@txt@toks
        \global\table@txtfalse
      \fi
    }%
    \hss
  }%
  \ifsinglecol
    \if@here \midinsert\global\@herefalse\else \topinsert\fi
      \box\TEMPBOX
    \endinsert
  \else
    \global\setbox\ItemBOX=\box\TEMPBOX
    \JoinStack
    \Print{Processing source for table \the\ItemNUMBER}%
  \fi
}

\def\UnloadZoneA{%
\FirstZoneAtrue
 \Iteration=\z@
  \loop
   \ifnum\Iteration<\LengthOfStack
    \GetItemSTATUS{\Iteration}
    \ifnum\ItemSTATUS=\InZoneA
     \GetItemBOX{\Iteration}
     \ifFirstZoneA \vbox to \BodgeHeight{\vfil}%
     \FirstZoneAfalse\fi
     \unvbox\ItemBOX\ItemSep
     \LeaveStack{\Iteration}
     \else
     \advance\Iteration \@ne
   \fi
 \repeat
}

\def\UnloadZoneC{%
\Iteration=\z@
  \loop
   \ifnum\Iteration<\LengthOfStack
    \GetItemSTATUS{\Iteration}
    \ifnum\ItemSTATUS=\InZoneC
     \GetItemBOX{\Iteration}
     \ItemSep\unvbox\ItemBOX
     \LeaveStack{\Iteration}
     \else
     \advance\Iteration \@ne
   \fi
 \repeat
}


\def\ShowItem#1{
  {\GetItemAll{#1}
  \Print{\the#1:
  {TYPE=\ifnum\ItemTYPE=\Figure Figure\else Table\fi}
  {NUMBER=\the\ItemNUMBER}
  {SPAN=\ifnum\ItemSPAN=\Single Single\else Double\fi}
  {SIZE=\the\ItemSIZE}}}
}

\def\ShowStack{%
 \Print{}
 \Print{LengthOfStack = \the\LengthOfStack}
 \ifnum\LengthOfStack=\z@ \Print{Stack is empty}\fi
 \Iteration=\z@
 \loop
 \ifnum\Iteration<\LengthOfStack
  \ShowItem{\Iteration}
  \advance\Iteration \@ne
 \repeat
}

\def\B#1#2{%
\hbox{\vrule\kern-0.4pt\vbox to #2{%
\hrule width #1\vfill\hrule}\kern-0.4pt\vrule}
}


\newif\ifsinglecol   \singlecolfalse

\def\onecolumn{%
  \global\output={\singlecoloutput}%
  \global\hsize=\PageWidth
  \global\vsize=\PageHeight
  \global\ColumnWidth=\hsize
  \global\TextLeading=12pt
  \global\Leading=12
  \global\singlecoltrue
  \global\let\onecolumn=\relax
  \global\let\footnote=\sing@footnote
  \global\let\vfootnote=\sing@vfootnote
  \ninepoint 
  \message{(Single column)}%
}

\def\singlecoloutput{%
  \shipout\vbox{\PageHead\pagebody\PageFoot}%
  \advancepageno
  \ifplate@page
    \shipout\vbox{%
      \sp@pagetrue
      \def\sp@type{plate}%
      \global\plate@pagefalse
      \PageHead\vbox to \PageHeight{\unvbox\plt@box\vfil}\PageFoot%
    }%
    \message{[plate]}%
    \advancepageno
  \fi
  \ifnum\outputpenalty>-\@MM \else\dosupereject\fi%
}

\def\ItemSep{\vskip\ItemSepamount\relax}

\def\ItemSepbreak{\par\ifdim\lastskip<\ItemSepamount
  \removelastskip\penalty-200\ItemSep\fi%
}


\let\@@endinsert=\endinsert 

\def\endinsert{\egroup 
  \if@mid \dimen@\ht\z@ \advance\dimen@\dp\z@ \advance\dimen@12\p@
    \advance\dimen@\pagetotal \advance\dimen@-\pageshrink
    \ifdim\dimen@>\pagegoal\@midfalse\p@gefalse\fi\fi
  \if@mid \ItemSep\box\z@\ItemSepbreak
  \else\insert\topins{\penalty100 
    \splittopskip\z@skip
    \splitmaxdepth\maxdimen \floatingpenalty\z@
    \ifp@ge \dimen@\dp\z@
    \vbox to\vsize{\unvbox\z@\kern-\dimen@}
    \else \box\z@\nobreak\ItemSep\fi}\fi\endgroup%
}


\def\gobbleone#1{}
\def\gobbletwo#1#2{}
\let\footnote=\gobbletwo 
\let\vfootnote=\gobbleone

\def\sing@footnote#1{\let\@sf\empty 
  \ifhmode\edef\@sf{\spacefactor\the\spacefactor}\/\fi
  \hbox{$^{\hbox{\eightpoint #1}}$}\@sf\sing@vfootnote{#1}%
}

\def\sing@vfootnote#1{\insert\footins\bgroup\eightpoint\b@ls{9pt}%
  \interlinepenalty\interfootnotelinepenalty
  \splittopskip\ht\strutbox 
  \splitmaxdepth\dp\strutbox \floatingpenalty\@MM
  \leftskip\z@skip \rightskip\z@skip \spaceskip\z@skip \xspaceskip\z@skip
  \noindent $^{\scriptstyle\hbox{#1}}$\hskip 4pt%
    \footstrut\futurelet\next\fo@t%
}

\def\footnoterule{\kern-3\p@ \hrule height \z@ \kern 3\p@}

\skip\footins=19.5pt plus 12pt minus 1pt
\count\footins=1000
\dimen\footins=\maxdimen


\def\landscape{%
  \global\TEMPDIMEN=\PageWidth
  \global\PageWidth=\PageHeight
  \global\PageHeight=\TEMPDIMEN
  \global\let\landscape=\relax
  \onecolumn
  \message{(landscape)}%
  \raggedbottom
}


\output{%
  \ifLeftCOL
    \global\setbox\LeftBOX=\vbox to \ZoneBSize{\box255\unvbox\ZoneBBOX}%
    \global\LeftCOLfalse
    \MakeRightCol
  \else
    \setbox\RightBOX=\vbox to \ZoneBSize{\box255\unvbox\ZoneBBOX}%
    \setbox\MidBOX=\hbox{\box\LeftBOX\hskip\ColumnGap\box\RightBOX}%
    \setbox\PageBOX=\vbox to \PageHeight{%
      \UnloadZoneA\box\MidBOX\UnloadZoneC}%
    \shipout\vbox{\PageHead\box\PageBOX\PageFoot}%
    \advancepageno
    \ifplate@page
      \shipout\vbox{%
        \sp@pagetrue
        \def\sp@type{plate}%
        \global\plate@pagefalse
        \PageHead\vbox to \PageHeight{\unvbox\plt@box\vfil}\PageFoot%
      }%
      \message{[plate]}%
      \advancepageno
    \fi
    \global\LeftCOLtrue
    \CleanStack
    \MakePage
  \fi
}


\Warn{\start@mess}

\def\mnmacrosloaded{} 

\catcode `\@=12 


\fi
\input psfig.sty


\def\eg{{e.g.~}}
\def\ie{{i.e.~}}
\def\etal{et~al.~}
\def\eV{e\kern-.15em V}                 
\def\keV{ke\kern-.15em V}                

\def\oiiia{[{\sc oiii}]$\lambda$5007}
\def\oiii{[{\sc oiii}]}
\def\oiiib{[{\sc oiii}]$\lambda$4959}
\def\oiiiab{[{\sc oiii}]$\lambda\lambda$4959,5007}
\def\niia{[{\sc nii}]$\lambda$6584}

\def\Ha{{\sc h}$\alpha$}
\def\Hb{{\sc h}$\beta$}

\def\feii{Fe{\sc ii}}

\def\tbrem{$T_{\rm brem}$}
\def\nh{$N_{\rm H}$}
\def\kms{km~s$^{-1}$}
\def\nhgal{$N_{\rm HGal}$}
\def\nhcm{$\times10^{20}$~cm$^{-2}$}

\def\rxj{RX~J1042+1212}

\def\aox{$\alpha_{\rm ox}$}

\def\ax{$\alpha_{\rm x}$}
\def\aopt{$\alpha_{\rm opt}$}
\def\lhard{$L_{\rm 2keV}$}

\def\lopt{$L_{5000}$}


\pageoffset{-2.5pc}{0pc}

%
  
%
%
%

\Autonumber  


\pagerange{000--000}    
\pubyear{0000}
\volume{000}

\begintopmatter  

\title{Double-peaked Balmer line emission in the radio-quiet AGN \rxj}

\author{E. M. Puchnarewicz, K.~O.~Mason and F.~J.~Carrera}

\affiliation{Mullard Space Science  Laboratory, University College London,
Holmbury St. Mary, Dorking, Surrey RH5 6NT, UK}

\shortauthor{E. M. Puchnarewicz, K. O. Mason and F. J. Carrera}
\shorttitle{Double-peaked Balmer lines in \rxj}



\abstract {We present optical and X-ray spectra of a radio-quiet X-ray selected
AGN, \rxj\ ($z$=0.271). The \Ha\ and \Hb\ emission lines are very broad (with
full widths at half maximum of $\sim$10000~\kms) and have double-peaked
profiles. Such  features are rarely observed in AGN in general but are even
more unusual in radio-quiet objects. The analysis of the {\sl ROSAT} PSPC data
reveals a non-varying, unabsorbed spectrum with an energy spectral index,
\ax=1.2 and little or no emission from a soft X-ray excess.  The slope of the
optical spectrum is similar, \aopt=1.0, and is consistent with an extrapolation
of the X-ray spectrum, suggesting that the same power-law continuum may
dominate throughout and that the big blue bump component is relatively weak. 

We look for a link between these various properties and investigate models of
double-peaked Balmer line emission in AGN. An accretion disc origin  is
unlikely in \rxj\ as this model predicts that lines emitted by a disc should
have a net gravitational redshift (both \Ha\ and \Hb\ have a net blueshift).
Emission from two broad line regions, each gravitationally bound to one
component of a supermassive black hole binary, is a possibility if the two
components are similar in size and nature. Alternatively, the lines (or at
least the narrow peaks of the lines) may be produced by a double-sided jet or
bipolar flow.}

\keywords{Galaxies: Seyfert -- Galaxies: Active -- X-rays: Galaxies -- Line:
profiles -- Galaxies: individual: \rxj.}

\maketitle  

\section{Introduction}

The structure and strength of the line emission in AGN reveals much about the
nature of the central engine and the broad line regions (BLRs). Double-peaked
Balmer line profiles are rarely observed but are of special interest because
they place specific constraints on the geometry and motions of the gas. Their
rarity is also intriguing; only $\sim$25-30 AGN with double-peaked profiles
have been observed and most of these ($\sim$80 per cent) are radio-loud, even
though radio-loud objects make up only $\sim$10 per cent of the overall AGN
population. Any successful model of double-peaked line-emitting AGN must also
explain why these objects prefer a radio-loud host.

Several models have been proposed as the origin of double-peaked lines, for
example the outer regions of an accretion disc (AD; \eg Collin-Souffrin \etal
1980; P\'erez \etal 1988; Alloin, Boisson \& Pelat 1988), from a binary BLR,
\ie two BLRs in orbit, each gravitationally bound to its own supermassive black
hole (Begelman, Blandford \& Rees 1980; Gaskell 1983,1988; Stockton \& Farnham
1991) or from a double-sided jet or cone (\eg Zheng, Binette \& Sulentic 1990;
Veilleux \& Zheng 1991; Zheng, Veilleux \& Grandi 1991).  For one model to be
generic to all AGN it must accommodate the wide range of properties observed,
which include  both red- and blue-dominant peaks and profile and flux
variability [including changes from red to blue dominant; see \eg Eracleous \&
Halpern (1994) for a sample of objects].

\begintable*{1}
\caption{{\bf Table 1.} X-ray and optical positions }
\halign{#\hfil &\quad\hfil#\hfil\quad   
               &\quad\hfil#\hfil\quad   
               &\quad\hfil#\hfil\quad   
               &\quad\hfil#\hfil\quad   \cr
                      & RA (J2000.0)
                      & Dec (J2000.0)
                      & V magnitude
                      & count rate \cr
 & & & & count s$^{-1}$\cr
\noalign{\medskip}
Optical$^1$ & 10 42 25.7 & +12 12 40 & 18.4 \cr
\noalign{\smallskip}
X-ray$^2$ (PSPC) & 10 42 25.9 & +12 12 46 &  & 4.7$\pm0.3\times10^{-2}$ \cr
}

\tabletext{$^1$From Mason \etal 1996;  the magnitude is measured from the
optical spectrum and should be considered uncertain, the positional error is
$\pm$1 arcsec. $^2$The count rate  is measured over 0.1-2.0~\keV\ and the
position is calculated from the centroid of the PSPC image;  positional errors
are a few arcseconds.}

\endtable 

The AD is particularly attractive as it is already a popular model for the
origin of the big blue bump (BBB; this component dominates the spectra of most
radio-quiet AGN, it rises through the optical/UV and the soft X-ray excess is
believed to be its high energy tail,  see \eg Edelson \&\ Malkan  1986; Walter
\& Fink 1993; Elvis \etal 1994).  However, circular ADs cannot predict
double-peaked  line profiles with dominant red peaks, although  Eracleous
\etal\ (1995) demonstrated that these problems and others related to profile
variability, could be overcome by invoking elliptical discs. All AD models
however predict that the line will have a net gravitational redshift (Eracleous
\etal 1995) yet some objects have lines which show a net blueshift, \eg
IC~4329A, Akn~120 and M81 (Marziani, Calvani \& Sulentic 1992; Bower \etal
1996; see also Eracleous \& Halpern 1994).  

Binary BLRs have been proposed as the source of double-peaked emission in
OX~169 (Stockton \& Farnham 1991) and possibly in  3C~390.3 (Zheng, Veilleux \&
Grandi 1991). However this model is not appropriate in all cases, for example 
Halpern \& Filippenko (1988) demonstrated that the positions of the lines in
Arp~120 did not vary within the timescale predicted. Peterson \etal (1990)
suggested that  the profiles observed in NGC~5548 may be due to a combination
of emission from both an AD {\sl and} a binary BLR. 

The double-sided jet model has the greatest degree of flexibility, due to the
large number of free parameters in the model and the possibility of different
conditions in the opposing jets, and was successfully used to fit the \Ha\
profile of 3C~390.3 (Zheng \etal 1991). Also, Storchi-Bergmann, Baldwin \&
Wilson (1993) suggested that the recent appearance of a broad, double-peaked
component in the LINER galaxy NGC~1097 (which is already known to contain four
faint optical jets) may be due to a new jet in formation. 

We present optical and X-ray data for a newly identified X-ray selected
radio-quiet AGN with double-peaked Balmer line emission. \rxj\ is a $z$=0.271
Seyfert~1 galaxy which was discovered during the optical identification
campaign of the {\sl ROSAT} International X-ray and Optical Survey (RIXOS;
Mason \etal 1996). Its \Ha\ and \Hb\ lines are strong and very broad
(FWHM$\sim$10000~\kms) and exhibit narrow peaks separated by $\sim$3000~\kms.
As a rare example of its kind, it presents a further opportunity to investigate
the nature of double-peaked emission in radio-quiet AGN. 

The optical and X-ray spectra of \rxj, including model fits to the soft X-ray
data, are presented in Section 2. In Section 3, we parameterize the emission
line properties by fitting Gaussian components to the \Ha\ and \Hb\ profiles.
We also investigate the continuum overall, from the radio to X-rays, and
compare these results and the emission line properties with those of other
samples of AGN. Our interpretation of these results in the context of current
models for double-peaked Balmer emission is discussed in Section 4.

\section{Observations}

\subsection{Optical}

\rxj\ was observed on 1994 February 16 using the Faint Object Spectrograph
(FOS) on the 2.9-m Isaac Newton Telescope at La Palma. The spectrum  covers a
range of 3500\AA\ to 10000\AA\ with a resolution of  15-20\AA\ FWHM in the red
and 8-10\AA\ FWHM in the blue. It was taken with a narrow slit positioned at
the parallactic angle and the exposure time was 600~s. 

\beginfigure*{1}
\psfig{figure=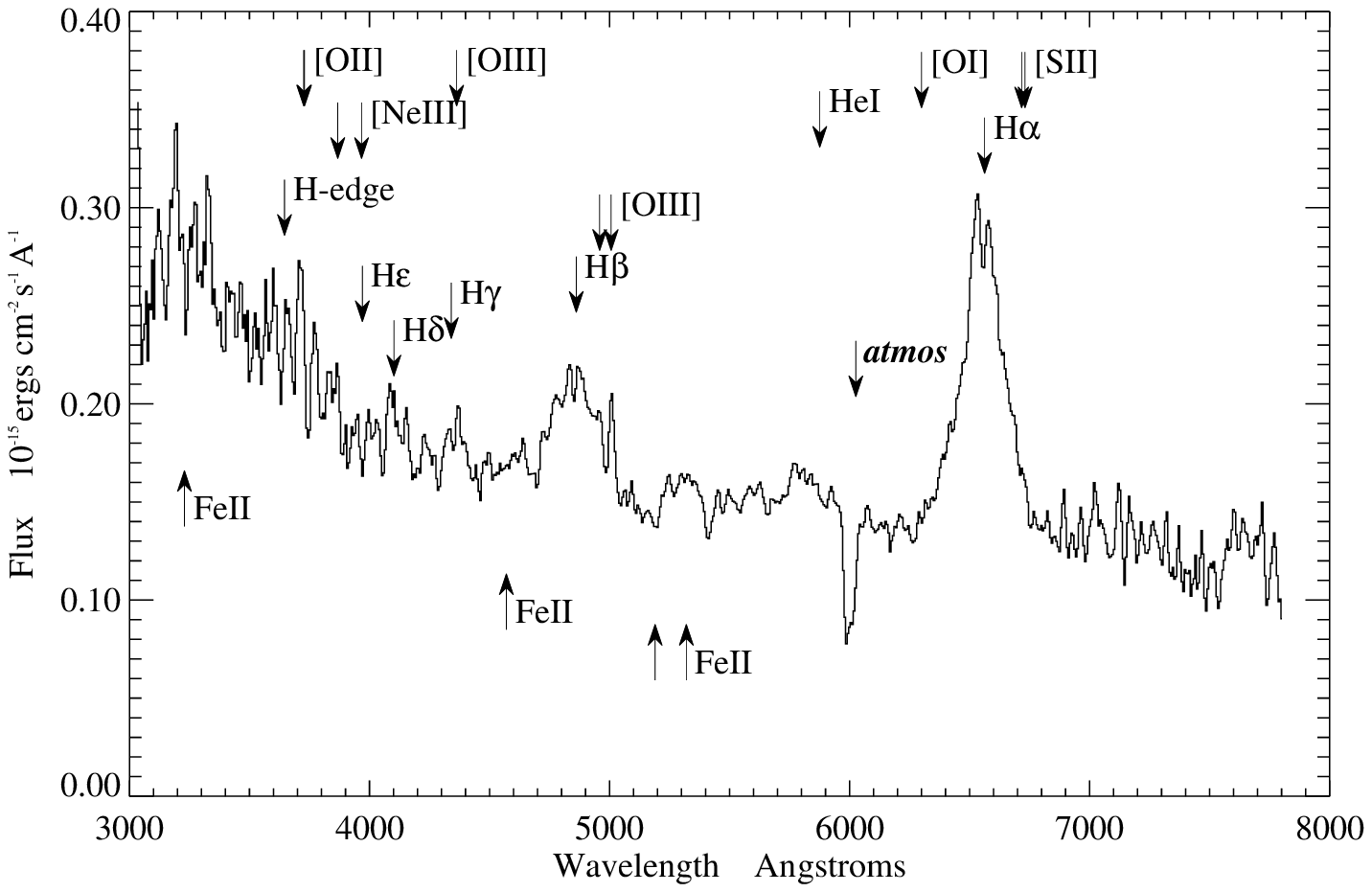,height=4.0in,width=6.0in,angle=0}
\caption{{\bf Figure 1.} The optical spectrum of \rxj\ taken at the INT in
1994 February. The spectrum has been smoothed using a boxcar filter with a
width of three bins and has been redshifted into the rest-frame of the AGN
(z=0.271). The expected positions of lines commonly found in AGN spectra are
indicated; the feature at $\sim$6000~\AA\ is caused by atmospheric extinction.}
\endfigure

The spectra were extracted and sky-subtracted using Mukai's (1990)
implementation of Horne's (1986) optimal extraction algorithm. Wavelength
calibrations were derived from Cu-Ar arc spectra while observations of
photometric standards were used for flux calibration. The spectrum is shown in
Fig~1; the optical position and V magnitude are listed in Table~1.

\subsection{Soft X-ray spectrum}

\subsubsection{Data}

The soft X-ray spectrum of \rxj\ was obtained on 1991 November 22 using the
Position Sensitive Proportional Counter (PSPC; Pfefferman \etal 1986)  on {\sl
ROSAT}; the exposure time was 10209~s. The PSPC has a 2$^\circ$ field of view
and covers a range of 0.1-2.4~\keV. Using the standard {\sc asterix} software,
the source spectrum was extracted using a circle with  radius 2.5 arcmin and a
source-free, annular region (with radii of 3.0 and 5.5 arcmin)  surrounding the
target was used for background subtraction. Counts were binned to yield at
least 20 per energy bin, ignoring channels 1-11 and 201-256 where the
response is uncertain. The spectrum has been corrected for all instrumental
effects including vignetting, dead-time and particle contamination and is
illustrated in Figure~2. The centroid of the X-ray emission, determined from
the PSPC image, is consistent with the optical position; the soft X-ray count
rate and position are listed in Table~1.  

\beginfigure{2}
\psfig{figure=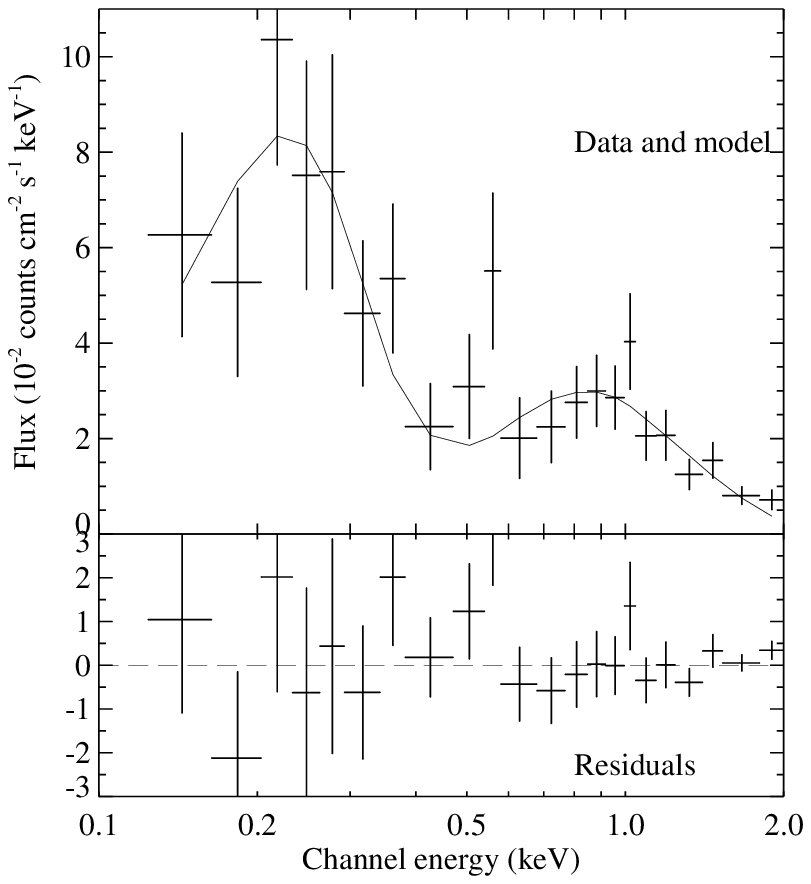,height=3.5in,width=3.3in,angle=0}
\caption{{\bf Figure 2.} [top] The {\sl ROSAT} PSPC spectrum of \rxj\ with the
best-fitting power-law model (thin line). [bottom] The residuals on the fit.}
\endfigure

\subsubsection{Model fitting}

The reduced PSPC data were fitted using the {\sc xspec} spectral fitting
software. The Galactic column in the direction of the source, \nhgal=2.8\nhcm,
was calculated by interpolating between the 21~cm measurements of Stark \etal
(1992). A single power-law with cold absorption provided a good fit to the data
($\chi_\nu^2$=0.9; see Fig 2), converging to an X-ray spectral energy index,
\ax, of 1.0$\pm0.7$ (errors are 90 per cent; \ax\ and all spectral indices are
defined such that $F_\nu\propto\nu^{-\alpha}$) and a Galactic column of
2.1$^{+2.2}_{-1.5}$\nhcm. Confidence contours for the power-law model are shown
in Fig~3 and show that the data are consistent with little or no absorption
intrinsic to the AGN (assuming of course that the power-law model is an
accurate representation of the intrinsic spectrum). 

\begintable*{2}
\caption{{\bf Table 2} Results of single power-law fits to the PSPC spectrum}
\halign{#\hfil &\quad\hfil#\hfil\quad   
               &\quad\hfil#\hfil\quad   
               &\quad\hfil#\hfil\quad   
               &\quad\hfil#\hfil\quad   \cr
Fit                   & Index 
                      & Normalization 
                      & Galactic \nh\ (\nhgal)
                      & $\chi_\nu^2$/dof \cr

                      & \ax\
                      &  \keV\ cm$^{-2}$ s$^{-1}$ \keV$^{-1}$ 
                      & 10$^{20}$ cm$^{-2}$
                      & K    \cr
\noalign{\medskip}
(1)                   & 1.0$\pm0.7$
                      & 1.5$\pm0.3\times10^{-4}$
                      & 2.1$^{+2.2}_{-1.5}$
                      & 17.9/19 \cr
\noalign{\smallskip}
(2)                   & 1.20$^{+0.16}_{-0.17}$   
                      & 1.6$\pm0.2\times10^{-4}$
                      & 2.8 (fixed)
                      & 18.5/20 \cr
}

\tabletext{(1) Best-fitting power-law  model  with the Galactic column (\nhgal)
free - errors are 90 per cent. (2) Best-fitting power-law model with \nhgal\
fixed  at the Stark \etal\ (1992) value.}

\endtable 

\beginfigure{3}
\psfig{figure=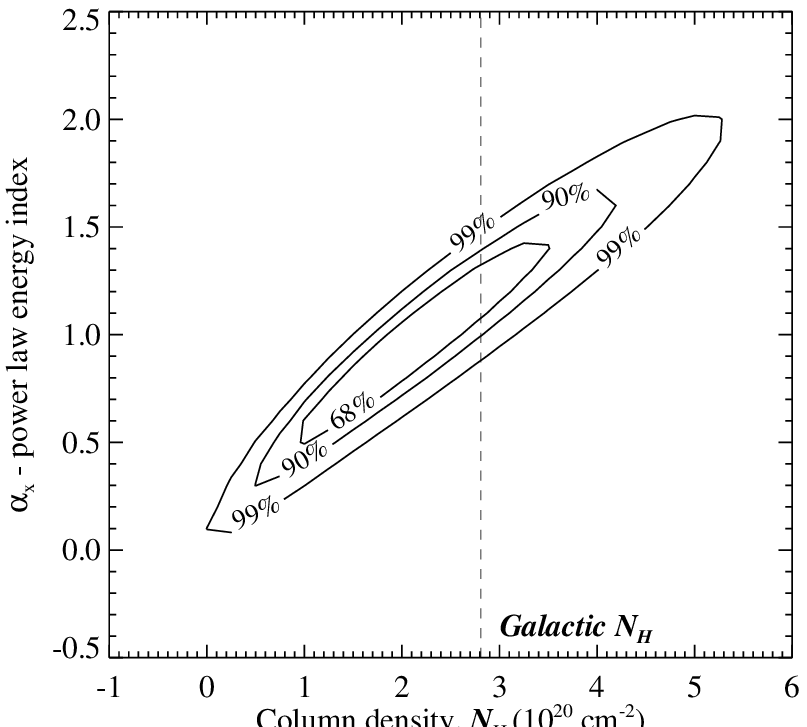,height=3.0in,width=3.3in,angle=0}
\caption{{\bf Figure 3.} Confidence contours for the power-law model fit to
\rxj. Also plotted as a thin, dashed line is the Galactic absorbing column
density interpolated from the Stark \etal\ (1992) measurements.}
\endfigure

If the absorbing column is fixed at the Galactic value, the best-fitting
power-law index converges to 1.20$^{+0.16}_{-0.17}$, with no significant change
to $\chi_\nu^2$ (=0.9; see Table 2 for details of all the X-ray spectral fits).
This model is similar to that used to fit the three-colour data in the original
RIXOS survey and gives a result which agrees within 90 per cent [\ie
0.96$^{+0.11}_{-0.11}$ (Mittaz \etal\ 1996); both sets of errors are quoted to
90 per cent].

\subsubsection{Variability}

To check for X-ray variability, we divided the PSPC data into the individual
observing intervals (OBIs; these comprised 10 periods of $\sim$1000~s taken
over two and a half days) and measured the count rate and spectrum for each.
However, we could find no evidence for significant flux or spectral variability
for the duration of the X-ray observation.

\section{The spectrum of \rxj}

\rxj\ was included in studies of the optical and X-ray properties of the RIXOS 
AGN (Puchnarewicz \etal 1996, hereafter P96; Puchnarewicz \etal 1997) and was
notable as an outlier from the anti-correlation between \ax\ and the Balmer
line FWHM, indeed \rxj\ has the broadest Balmer emission of all AGN in the
RIXOS sample. In the following analysis, we present an investigation of the
emission line and continuum properties of \rxj.


\subsection{Double-peaked Balmer lines}

A closer examination of the optical spectrum (plotted in Fig 1) reveals that
both \Ha\ and \Hb\ are double-peaked as well as being very broad (see also Figs
4-6). In order to model the complex shapes of the Balmer line profiles and to
measure line fluxes and EWs, we have fitted the data with combinations of
Gaussians (using a second-order polynomial for the underlying continuum). 

\subsubsection{\Ha}

Initially, the \Ha\  feature was fitted with a single component for the line
emission and an ``absorption dip'' (see Fig 4a). The FWHM of the broad emission
component is 8900 km s$^{-1}$ (the FWHM in Tables 3 and 4 have been deconvolved
from the instrumental profile which is $\sim$800~\kms) and both the ``dip'' and
the line have a central wavelength of 6553~\AA. 
A better representation of the data was obtained using a three-component fit,
with a broad underlying component and separate red and blue peaks (Fig 4b). For
this model, the FWHM of the broad component is 10900 km s$^{-1}$ while the red
and blue peaks are much narrower (1500 km s$^{-1}$ and 1400 km s$^{-1}$
respectively). The blue peak is slightly stronger than the red and the
separation between the two is 2600~\kms. 

The fits to \Ha\ are illustrated in Fig 4 and detailed in Table 3.

\beginfigure{4}
\psfig{figure=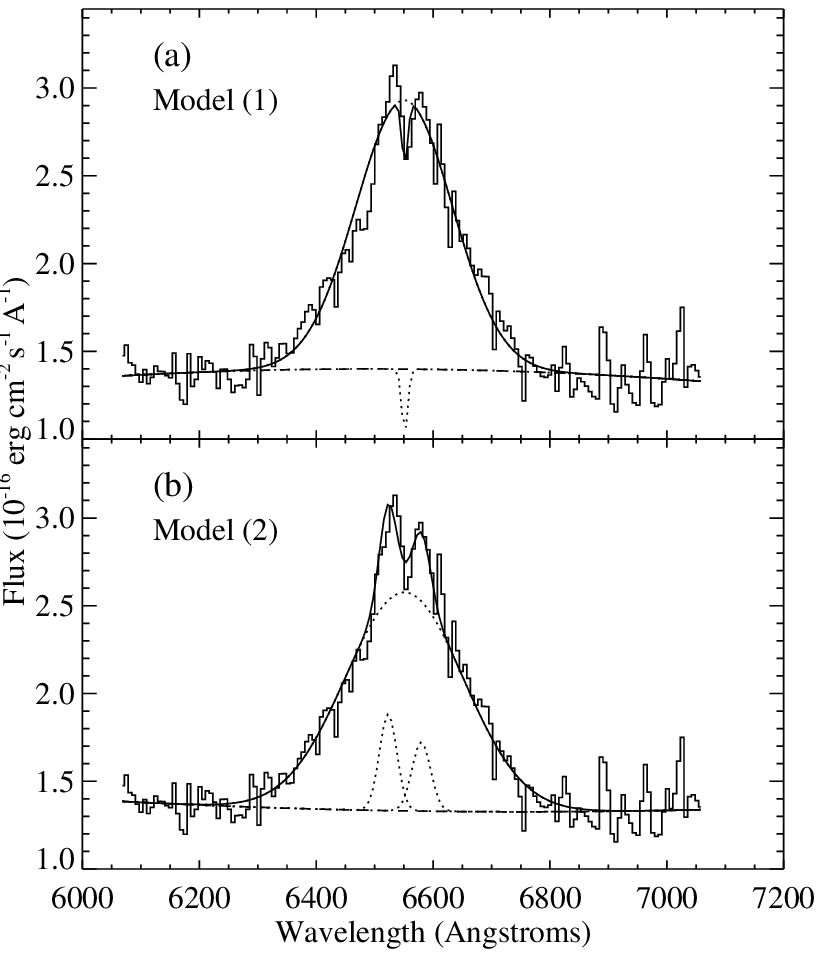,height=4.0in,width=3.3in,angle=0}
\caption{{\bf Figure 4.} The profile of \Ha\ (histogram) compared with 
multi-Gaussian fits to the data: {\sl (a)} a broad emission feature plus an
``absorption dip'' [model (1) from Table 3];  {\sl (b)} a broad underlying
component plus red and blue peaks [model (2) from Table 3]. The total fit is
plotted as a solid line, the fit to the continuum as a dashed line and fits to
individual gaussian components (plus the continuum) as dotted lines.}
\endfigure

\begintable*{3} 
\caption{{\bf Table 3.} Measurements of the \Ha\ parameters}
\halign{     \hfil#\hfil\quad
            &\quad#\hfil\quad   
       &\quad\hfil#\hfil\quad   
       &\quad\hfil#\quad    
       &\quad\hfil#\hfil\quad   
       &\quad\hfil#\quad   \cr 
Model                & Component  & Position & FWHM       & Flux  & EW  \cr 
                     &            & \AA\     & km s$^{-1}$ 
                     & 10$^{-14}$ erg cm$^{-2}$ s$^{-1}$ & \AA\hfil \cr 
\noalign{\medskip}
(1)                  & emission line & 6553     &  8900  &   3.3  & 240    \cr 
                     & absorption dip & 6553    & unresolved & --0.05 & --3 \cr
\noalign{\smallskip}
 (2)                 & broad      & 6553  & 10900  & 2.9 & 220 \cr
                     & blue peak  & 6526  &  1400  & 0.2 &  20 \cr
                     & red peak   & 6583  &  1500  & 0.2 &  10 \cr 
}                                                               
\tabletext{(1) Two-component model with one Gaussian representing the \Ha\
emission line and a second the ``absorption dip''. (2) Three-Gaussian model
with a very broad underlying component and red and blue intermediate-width
peaks. Measurements are taken from the spectrum shown in Figure~1. The position
is the wavelength of the Gaussian component; the FWHM is calculated from the
Gaussian $\sigma$ and has been deconvolved from the instrumental profile. The
flux and EW have been measured directly from the spectrum after the continuum
and model flux from other fitted components have been subtracted.} 
\endtable 

\subsubsection{\Hb}

Fitting the \Hb\ profile is less straightforward (see Fig 5). It  is likely to
be blended with optical \feii\ as this is very strong elsewhere in this region
(see the blends at 4570~\AA\ and 5250~\AA\ in Fig 1). By fitting a single
Gaussian to \Hb\ (and narrower components to \oiiiab; see Fig 5a), we derive a
FWHM (for \Hb) of 11300 km s$^{-1}$. We also tried adding an absorption dip in
\Hb\ to this model (Fig 5b), although this made little difference to the
overall fits (which may be largely due to the relatively poor data quality
around \Hb). Like \Ha, the dip is slightly blueshifted with respect to \oiiiab.
The \oiiiab\ lines are unresolved and their flux is low compared to \Ha\ and
\Hb. Parameters from these fits are given in Table 4.  

\beginfigure{5}
\psfig{figure=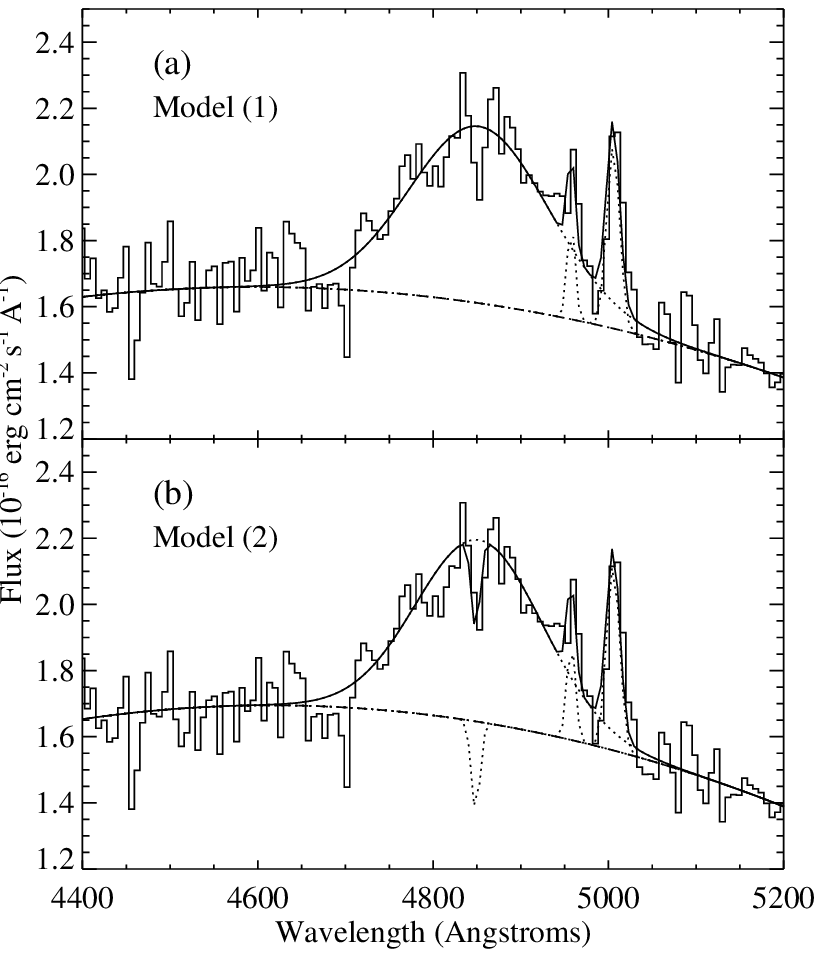,height=4.0in,width=3.3in,angle=0}
\caption{{\bf Figure 5.} The profile of \Hb\ (histogram) compared with 
multi-Gaussian fits to the data: {\sl (a)} a broad \Hb\ component plus narrower
components for \oiiiab; [model (1) in Table 4];   {\sl (b)} the same as (a)
but with an additional ``absorption dip'' in \Hb [model (2) in Table 4]. The
total fit is plotted as a solid line, the fit to the continuum as a dashed line
and fits to individual gaussian components (plus the continuum) as dotted
lines.}
\endfigure

\beginfigure*{6}
\psfig{figure=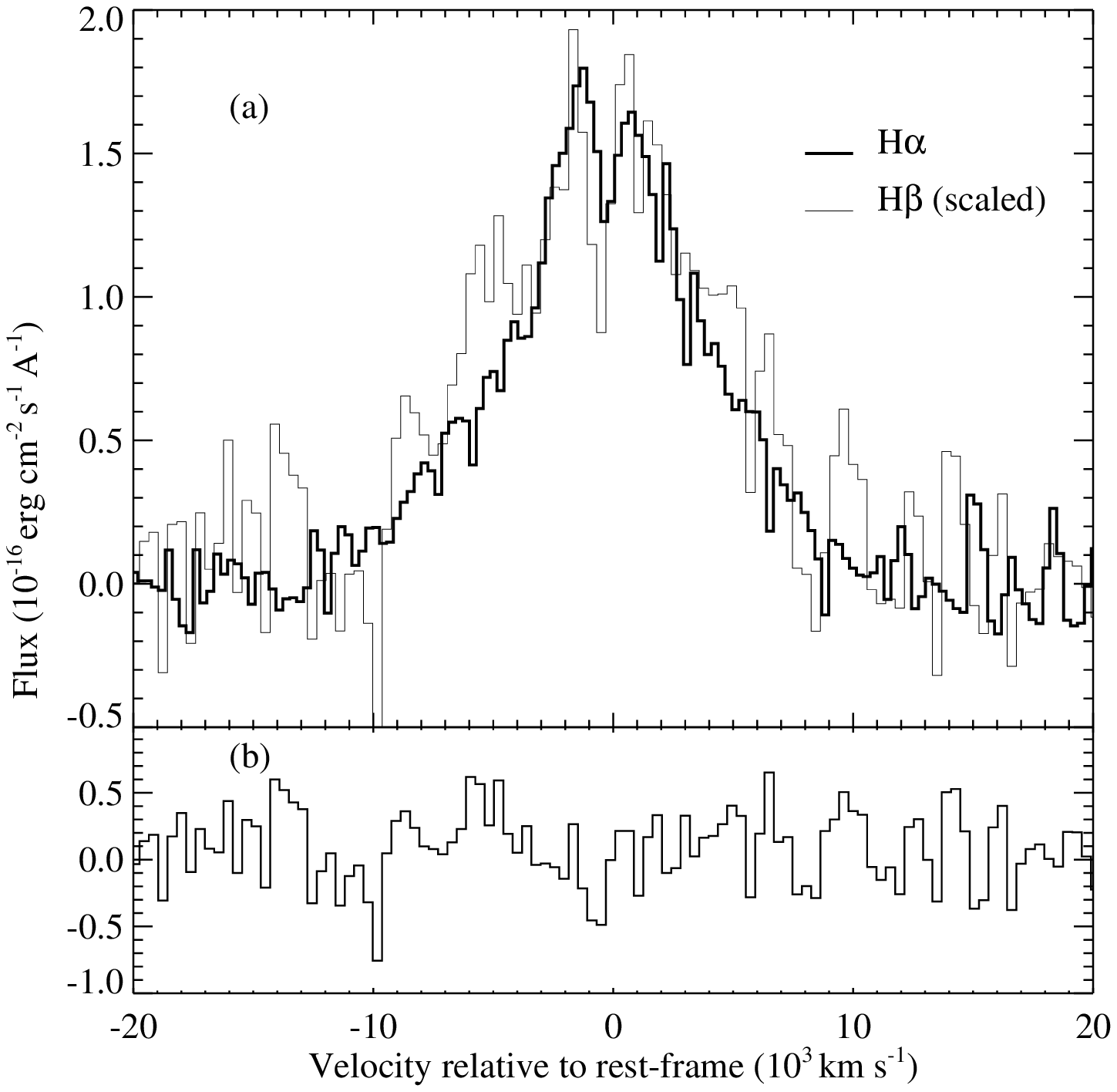,height=6.0in,width=6.0in,angle=0}
\caption{{\bf Figure 6.} {\sl (a)} A direct comparison of the \Ha\ and \Hb\ 
line profiles. \Ha\ is plotted as a thick histogram while \Hb\ is plotted
as a thin histogram. For \Ha, the fitted continuum from model (1) in Table 3 has
been subtracted. For \Hb, the fits to the continuum and to the \oiiiab\ lines 
using model (2) in Table 4 have been subtracted and the remaining data have
then been multiplied by a factor of 2.8 to match the \Ha\ profile. {\sl (b)}
The difference between the \Ha\ and scaled \Hb\ profiles (\ie \Hb-\Ha).}
\endfigure

\begintable*{4}
\caption{{\bf Table 4.} Measurements of \Hb\ and \oiiiab\ parameters.}
\halign{             \hfil#\hfil\quad   
                    &\quad#\hfil\quad   
               &\quad\hfil#\hfil\quad   
               &\quad\hfil#\quad   
               &\quad\hfil#\hfil\quad   
               &\quad\hfil#\quad   \cr
Model                & Component  & Position & FWHM        & Flux  & EW  \cr
                     &            & \AA\     & km s$^{-1}$ & 10$^{-14}$ erg cm$^{-2}$ s$^{-1}$ & \AA\hfil \cr
\noalign{\medskip}
(1)                  & \Hb\       & 4854     & 11300  &   0.72   & 50    \cr
                     & \oiiia\    & 5007     & unresolved &   0.08  &  6    \cr
                     & \oiiib\    & 4959     & unresolved &   0.02  &  2    \cr
\noalign{\smallskip}
(2)                  & \Hb\       & 4855     & 10400  &   0.67   & 40    \cr
                     & absorption dip & 4851 & unresolved & --0.04  &  1    \cr
                     & \oiiia\    & 5007     & unresolved &   0.09  &  7    \cr
                     & \oiiib\    & 4959     & unresolved &   0.03  &  2    \cr
}                                                               
\tabletext{(1) Three-component model with separate Gaussians representing the
\Hb, \oiiia\ and \oiiib\ lines. (2) Four-component model with an additional
``absorption dip'' in \Hb. Other details are the same as for Table 3.}
\endtable 

\subsubsection{Balmer line profiles}

In Fig 6, we compare the profiles of \Ha\ and \Hb\ directly. The fitted
continuum [from model (2) in Table 3]  has been subtracted from the region
around \Ha. For \Hb, the fitted continuum and fits to the \oiiiab\ lines [using
model (1) in Table 4] have been subtracted, then the remaining data have been
multiplied by a factor of 2.8 to match the \Ha\ peak. The profiles have a
similar shape and both display the double-peaked profile; the blue peak appears
to be stronger than the red in \Ha. The excess emission in the wings of \Hb\
relative to \Ha\ (see Fig 6) may be due to \feii.

The centre of the ``red peak'' of the \Ha\ line is coincident with the expected
position of \niia, however there are a number of reasons why we believe that
this is not caused by \niia\ emission. Firstly, despite the poorer signal in
the \Hb\ region, the \Ha\ and \Hb\ profiles show a remarkable similarity when
plotted in velocity space (see Fig 6); note in particular the relative
positions of the peaks and the dip which are very closely matched. This is
unlikely to be coincidental. Secondly, if the red peak was actually due to
\niia, this would imply {\sl (a)}  that the blue peak was a separate narrow
component of \Ha, blueshifted with respect to the other (forbidden) narrow
lines by $\sim$1600~\kms; and {\sl (b)} that there was no significant \Ha\
emission from the narrow line region (\ie at 6562~\AA), despite strong \niia\
emission at the systemic velocity (which in any case, is generally very weak in
luminous AGN like \rxj). A physical interpretation of this would be difficult.
Finally, the \oiiia\ line is unresolved therefore one would also expect the
same of any \niia\ emission, yet the red peak in \Ha\ has been resolved. 

The observed \Ha/\Hb\ flux ratio is 5$^{+1}_{-4}$  (these errors have been
estimated by fitting the upper and lower limits on the Gaussian profiles by
eye) and the ratio of their peaks is 2.8. Assuming that the intrinsic \Ha/\Hb\
flux ratio is 2.8 (the appropriate value for case B recombination), then there
is no  evidence for significant dust reddening in \rxj. In Section 2.2.2 we
showed that there was no evidence either for intrinsic absorption in cold gas,
for a single power-law model fit to the X-ray spectrum.

\subsection{Multiwavelength continuum}

The optical and X-ray spectra of \rxj\ in the rest-frame of the AGN  are
plotted in Fig 7. We have also searched in the radio and IR for any emission
associated with this source, however {\sl no} 20~cm  radio source was found
stronger than 1.4~mJy within a few arcminutes of the optical position of \rxj\
(J. J. Condon, private communication), neither are there any {\sl IRAS} sources
within a radius of 20~arcmin. We conclude that this is a radio-quiet AGN, with
no evidence for a strong IR dust component.

The data shown in Fig~7 suggest that the optical continuum may be dominated by
an extrapolation of the X-ray power-law and that the BBB and the soft X-ray
excess are relatively weak.  The optical spectrum shows no features typical of
significant emission from the host galaxy (see Fig 1). The apparent turn-up
towards the blue shortward of 4000~\AA\  may be the so-called `little' blue
bump (Wills, Netzer \& Wills 1985), a false excess continuum caused by the
combination of a strong Balmer continuum and FeII blends. Such a feature would
not be unexpected in \rxj\ given the strength of the Balmer {\sl lines} and
FeII at other wavelengths. 

\beginfigure{7}
\psfig{figure=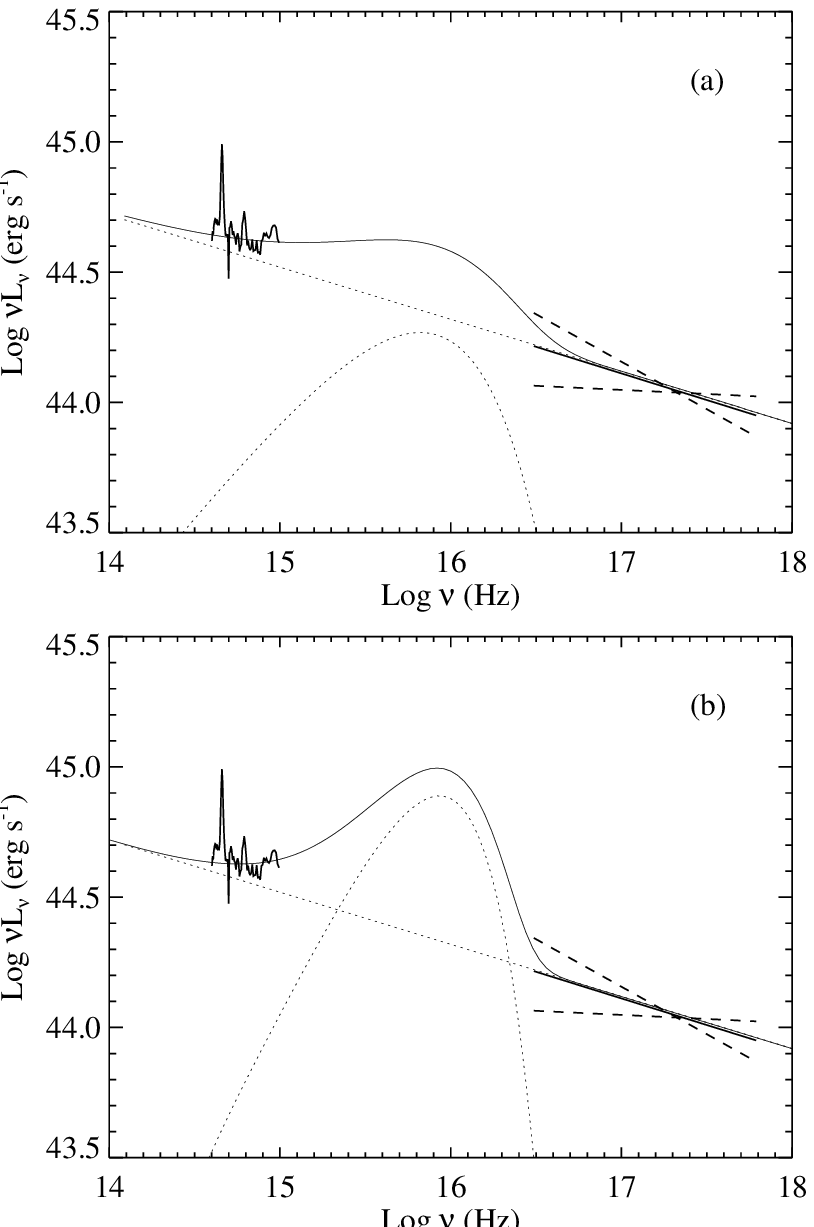,height=5in,width=3.3in,angle=0}
\caption{{\bf Figure 7.} The optical and X-ray spectra of \rxj\ are compared
here with two different models for the optical to X-ray continuum.  In {\sl
(a)}, the BBB is modeled using a bremsstrahlung spectrum with
\tbrem=4$\times10^5$~K, while in {\sl (b)}, a simple thin AD model with
$T$=25~\eV\ has been assumed.  In both cases there is an underlying
$\alpha$=1.2 power-law and different thermal  models have been added to assess
the an upper limit on the strength of the BBB. The sum of the thermal and
non-thermal spectra are plotted as solid lines and  individual components as
dotted lines. }
\endfigure

In most radio-quiet AGN, the BBB dominates the energy output of the source.
When this component is strong, the optical slope (\aopt) is very hard (\ie it
rises strongly towards the blue) and the optical to X-ray ratio (\aox) is high.
For example, in samples of optically-selected AGN (such samples preferentially
select objects with strong BBBs), the median \aopt\ is $\sim$0.2-0.3  ( \eg
Neugebauer \etal 1987; Francis \etal 1991)  while in the Wilkes \etal (1994)
sample of PG quasars, the mean \aox=1.50. In contrast, the RIXOS AGN sample,
which is X-ray selected, has a softer mean \aopt\ (=0.9) and a lower mean \aox\
(1.1), \ie the `mean' optical to X-ray continuum has no significant BBB
component and a spectral index, $\alpha\sim$1. The \aopt\ and \aox\ for \rxj\
are both typical of the means of the RIXOS AGN [\ie for \rxj, \aopt=1.0$\pm$0.5
(P96) and \aox=1.20$\pm$0.14], thus we find that the BBB is probably 
intrinsically weak in this source.  

Nonetheless, we have no data in the UV or the EUV where the peak of the BBB
generally lies and it is possible that the BBB may emerge significantly above
the underlying continuum out of our observed range. Therefore we have estimated
the upper limit  on the strength of a possible thermal UV component by
comparing  \rxj\ with two different models of the BBB and  using the optical
and X-ray data as constraints at low and high energies.  A thermal
bremsstrahlung spectrum and a simple AD model (Pringle 1981) were used to
represent the BBB (Fig~7a and b respectively), while in each case, the
best-fitting power-law  model to the X-ray data (model 2 in Table 2; \ax=1.2)
was used for the underlying non-thermal continuum. The bremsstrahlung spectrum
has a temperature, \tbrem=4$\times10^5$~K (in the AGN rest-frame) and a total
luminosity of 6$\times10^{44}$ erg s$^{-1}$ (all luminosities have been
calculated assuming a value of 50~\kms~Mpc$^{-1}$ for the Hubble constant,
$H_0$ and 0 for the deceleration parameter, $q_0$). The AD has a maximum
temperature of 25~\eV, a luminosity of 1.5$\times10^{45}$ erg s$^{-1}$ and the
mass of the black hole is 4$\times10^6 M_\odot$. The luminosity in the
power-law from log$\nu$=14 to 18~Hz, is 2.1$\times10^{45}$ erg s$^{-1}$.

By integrating these models over log$\nu$=14 to 18~Hz, we measure total
luminosities of $L_{\rm opt,X-ray}=3\times10^{45}$ and $4\times10^{45}$ erg
s$^{-1}$ for the bremsstrahlung and AD models respectively. If we assume  that
this part of the spectrum dominates the bolometric luminosity, $L_{\rm bol}$
(since \rxj\ is not detected longwards of 10$\mu$m),  this would imply a mass
for the central black hole of at least $\sim2\times10^7 M_\odot$ (from the
Eddington limit); for the AD model this in turn suggests that the system would
be super-Eddington, \ie $L\sim10L_{\rm Edd}$ (and thus the thin disc
approximation we have used is not self-consistent).

%
%

\subsection{Further comparisons with other AGN}

\subsubsection{Balmer line properties}

The mean \Ha\ FWHM for the RIXOS AGN is 2300$\pm$2200 \kms\ (1$\sigma$ standard
deviation) thus in \rxj, where the FWHM are $\sim$10000~\kms\ (see Tables 3 and
4), the Balmer lines are very broad for an X-ray selected object. They are also
broad when compared to optically-selected quasars, \eg for the Boroson \& Green
(1992; hereafter BG) sample, we calculate a mean \Hb\ FWHM of
3800$\pm$2100~\kms\ (1$\sigma$). They are closer to the range for the
``disklike'' emitters from the Eracleous \& Halpern (1994) sample of radio
galaxies and radio-loud quasars. These are objects whose Balmer line profiles
are double-peaked and are well-fitted by an accretion disk model; they have a
mean \Ha\ FWHM of 12500$\pm$4900~\kms\ (1$\sigma$) and a range of 6800 to
23200~\kms.

The EWs of the Balmer lines in \rxj\ are both high relative to the means for
the RIXOS sample which has means of 100~\AA\ and 40~\AA\ for \Ha\ and \Hb\
respectively.  However, the ratios of the Balmer line luminosities to the X-ray
luminosity (\lhard) {\sl are} typical of other objects in the RIXOS sample,
suggesting that the lines are merely responding to an enhanced ionizing
continuum. 

\subsubsection{\oiii}

The EW of \oiiia\ is $\sim$6~\AA\ which is weak relative to the means for the
RIXOS sample (30~\AA), the Stephens sample (44~\AA) and for the BG sample
(24~\AA). However the \oiiia\ luminosity (41.5 erg s$^{-1}$) is similar to the
means that we calculate for the RIXOS and Stephens (1989)  samples
(41.7$\pm$0.6 erg s$^{-1}$ and 41.9$\pm$0.6 erg s$^{-1}$ respectively),
suggesting that it is a high optical continuum which causes the low EW, rather
than intrinsically weak line emission. This may be due to variability, \eg if 
the nuclear spectrum has strengthened recently so that the narrow line region
has not yet had time to respond.

\subsubsection{Optical and X-ray continua}

We find that the log of the continuum luminosity at 5000~\AA,
\lopt=29.7$\pm$0.3, is somewhat high compared to the RIXOS AGN where the mean
\lopt=29.2$\pm$0.5 (1$\sigma$ standard deviation; this mean has been calculated
only for those sources whose $z$ is low enough to permit the detection of \Hb,
\ie at $z<0.95$). However if we compare the absolute V magnitude of \rxj\
($M_{\rm V}$=--22.4) with that of the BG UV-excess quasars, we find that it is
relatively weak in the optical  (we calculate a mean for the BG objects of
--24.2$\pm$1.5), yet the lines in \rxj\ are broader and double-peaked.
Apparently, it is not simply the strength of the optical continuum which forms
these Balmer lines.
                               
The optical to X-ray continuum in \rxj\ seems to be remarkably flat overall
(\ie \aopt=1.0$\pm$0.5, \aox=1.2$\pm0.1$ and \ax=1.2$\pm$0.2), suggesting that
a non-thermal component dominates throughout. It is possible then that the
presence of a strong power-law spectrum is related to the production of the
broad, double-peaked lines. We have thus searched the RIXOS AGN for objects
with similar spectra, \ie those whose \ax, \aopt\ and \aox\ was between 0.5 and
1.5,  whose \lopt\ was greater than 29.5 erg s$^{-1}$ and whose optical 
spectrum covered the \Hb\ region. There are two AGN (in addition to \rxj) which
meet these criteria,  source 301 in field 232 (F232-301) and source 30 in field
211 (F211-030). However, F215-019 and F232-301 have \Hb\ FWHM of only
1000~\kms\ and 2600~\kms\ respectively, thus it is unlikely that the line
profiles in \rxj\ are solely due to a strong non-thermal component.

\subsection{Summary}

We find that the \Ha\ and \Hb\ line emission in \rxj\ is very broad and
double-peaked. The BBB is relatively weak and the optical to X-ray continuum
may be dominated by a non-thermal component with a slope, $\alpha\sim$1.2. This
source was not detected at radio or IR wavelengths.

The double-peaked line emission cannot be exclusively related to the shape
and/or the strength of the ionizing continuum since there are examples of AGN
which do not have broad or double-peaked line profiles yet have optical to
X-ray continua which are otherwise similar in nature. This implies that there
may be some phenomenological difference in \rxj\ which has (as yet) only
manifested itself in the Balmer line emission, although it may have some
connection with the strong non-thermal component.

\section{Discussion}

The discovery of broad, double-peaked Balmer emission in \rxj\ is intriguing.
The double peaks indicate that for the low-velocity broad line-emitting gas  at
least, there are two distinct components, one moving towards us and the other
moving away (relative to the systemic velocity; this may also be true of the
high-velocity gas). Yet double-peaked  line emission is an unusual property of
AGN in general and it is found mostly in radio-{\sl loud} objects (see
Eracleous \& Halpern 1994 and references therein), even though these are
outnumbered by their radio-quiet counterparts in the general AGN population by
$\sim$10 to one. As a {\sl radio-quiet} double-peaked AGN \rxj\ is thus
extremely rare [other examples are NGC~5548 (Peterson, Korista \& Cota 1987),
Akn~120 (Peterson \etal 1983; Marziani, Calvani \& Sulentic 1992) and IC~4329A
(Marziani \etal 1992)], which adds to its already strange nature.

In this section, we discuss models of double-peaked line profiles to establish
whether they can also predict other properties of \rxj, such as the strong high
velocity BLR and the dominating power-law.

\subsection{Models for double-peaked emission}

\subsubsection{Accretion disks}

Optically-thick accretion disks (ADs) are often invoked to explain the origin
of the BBB in AGN (\eg Shakura \& Sunyaev 1973; Czerny \& Elvis 1987; Madau
1988; Sun \& Malkan 1989) and it has also been suggested that conditions at the
outer edges of an AD may be suitable for the production of the low-ionization
lines (which include \Ha\ and \Hb; Collin-Souffrin \etal 1980; Collin-Souffrin
1987). The presence of an additional  photoionizing source, \eg by a
non-thermal source, a hot corona or an ion torus (Rees \etal 1982) is required
to power the line emission, as the gravitational energy available in the disk
is barely sufficient on its own (\eg Dumont \& Collin-Souffrin 1990; Marziani
\etal 1992). The profiles of lines from an AD are expected to be double-peaked
(\eg Chen \etal 1989; Fabian \etal 1989; Dumont \& Collin-Souffrin 1990) and AD
models have been successfully used to predict the Balmer lines of several AGN.
For example, Eracleous \& Halpern (1994) surveyed the \Ha\ emission of 94 radio
galaxies and radio-loud quasars and identified 12 objects whose line profiles
were well-fitted by an AD. Other studies of individual objects include those of 
P\'erez \etal (1988; 3C~390.3), Alloin \etal\ (1988; Akn~120) and
Rodriguez-Ardila \etal (1996; C16.16). ADs have also been invoked to explain
the skewed Fe~K$\alpha$ line observed in the ASCA spectrum of the Seyfert 1
galaxy MCG-6-30-15 (Tanaka \etal 1995).

AD models make certain predictions about the shape of the emission lines. The
blue peak will always be stronger than the red peak, due to relativistic
beaming, and the entire line will have a net gravitational redshift (\eg Chen
\& Halpern 1989). Yet in about one-quarter of double-peaked  objects, the red
peak is stronger than the blue (\eg PKS~1020--103; Eracleous \& Halpern 1994)
while in 3C~390.3 for example, the \Ha\ profile is variable and has been
observed to change from blue-dominant to red-dominant (Zheng \etal 1991),
contradicting AD models. Some of these problems have been overcome by Eracleous
\etal\ (1995) who invoked elliptical disks, formed by the tidal perturbation of
a binary black hole or from the debris of a tidally disrupted star.

In the case of \rxj, we find that the blue peak of \Ha\ is slightly stronger
than the red (Fig 4 and Table 3) but there is no evidence for any net
gravitational redshift of the line; the central wavelength of the single
component fit lies at 6553~\AA, which would rather imply a net {\sl blueshift}
equivalent to a velocity of $\sim$500~\kms. This is inconsistent with the AD
model which, even in the elliptical case, predicts that the line will always
have a net redshift (Eracleous \& Halpern 1994). Thus we conclude that an AD is
unlikely to be the origin of the Balmer lines in \rxj.

\subsubsection{A binary BLR}

One alternative to the AD model is that of two separate BLRs orbiting a
supermassive black hole binary system (Gaskell 1983);  the existence of 
supermassive binaries was first discussed in detail by Begelman, Blandford \&
Rees (1980). In this model, the BLRs are gravitationally bound to their
respective black holes which are in orbit around each other, so that the line
profiles for each BLR are separated and displaced from the centre. It has also
been suggested that even separate orbiting  {\sl narrow-}line regions (NLRs)
exist in Mkn~78 and Mkn~266 (see Gaskell 1988). However, Chen \etal (1989) have
argued that the line profile which would actually be observed from a binary BLR
would rarely be double-peaked; the low-velocity clouds, which presumably lie at
large distances, would not orbit their respective black holes but the centre of
mass of the two, `filling-in' the core of the profile. However these clouds may
have been stripped away during the formation of the binary (Begelman \etal
1980) or the  cloud velocities may not be dominated by the gravitational
potential (Stockton \& Farnham 1991); in either case a double-peaked profile
may be observed.

While investigating their models of supermassive black hole binaries, Begelman
\etal (1980) envisaged that the ratio of the black hole masses would be 
$\sim$10. In this case, activity at the larger component is likely to 
dominate the system, so that just  a single displaced emission line would be
observed. However it is possible that two similar, independent nuclei may form
the binary and this was in fact  suggested for OX~169 by Stockton \& Farnham
(1991) and may also be true of \rxj, given that the Balmer line profiles are
similar for these two AGN.

A measurement of the ratio of the black hole masses in a supermassive binary 
may be derived from the  relative velocities of the red and blue peaks.  The
blue peak in \rxj\ has a velocity of --1600 km s$^{-1}$ relative to the
systemic velocity (assuming that this is defined by the redshift of \oiiia) and
the red peak has a relative velocity of 1000~\kms. This gives a mass ratio of
1.6, which is similar to that derived for OX~169 (2.9; Stockton \& Farnham
1991). A lower limit for the total mass of the black holes may be calculated
using the Eddington limit. We estimate a bolometric luminosity, $L_{\rm bol}$
of  $\sim4\times10^{45}$ erg s$^{-1}$ from the optical to X-ray fits shown in
Figure 7, which in turn implies a black hole mass, $M>2\times10^7 M_\odot$.                                                         

\beginfigure{8}
\psfig{figure=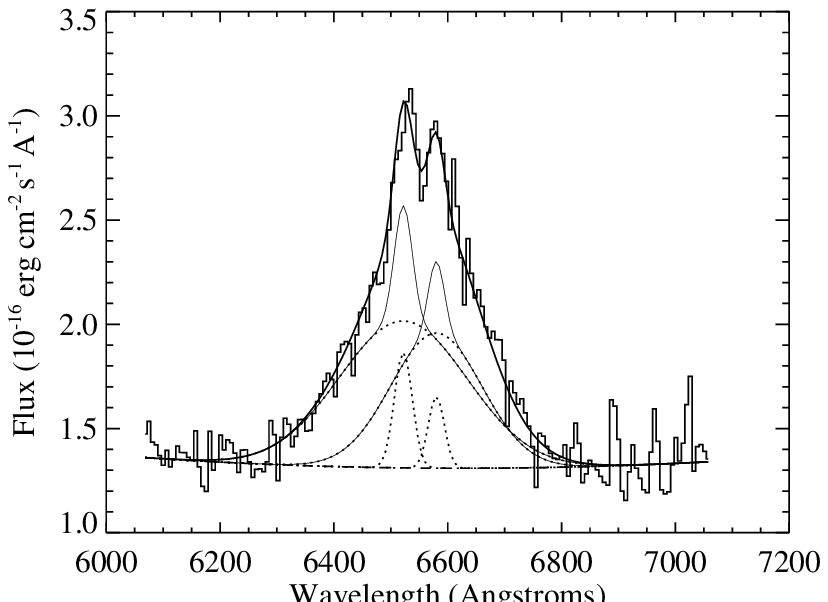,height=2.5in,width=3.3in,angle=0}
\caption{{\bf Figure 8.} The \Ha\ profile compared with a four-Gaussian
component fit. The total fit is plotted as a thick solid line, the continuum as
a dashed line and individual Gaussians (plus the continuum) as dotted lines.
The `pair' of profiles implied by the fits is plotted with thin, solid lines.}
\endfigure

The model for the \Ha\ profile shown in Fig~4b uses a single, broad underlying
Gaussian plus two, narrower peaks. Taken literally, this would suggest the
presence of one very broad line region  and two distinct ``intermediate'' line
regions which is clearly inconsistent with the binary BLR model. A more
appropriate model would have two high- and two low-velocity regions of the BLR,
so we have tested this by fitting four such Gaussians to the \Ha\ profile. The
positions of each of the Gaussian pairs were fixed at those for the red and
blue peaks in model (2) for \Ha, \ie at 6526~\AA\ and 6583~\AA, and all other
parameters were free. The best fit is shown in Fig 8 and compares well with the
data. The FWHM of the broad red and blue components are 12200~\kms\ and
8600~\kms\ respectively and their flux ratio is 1.5. The FWHM of the narrow
components are 1500~\kms\ and 1400~\kms\ for the red and blue components
respectively, while their flux ratio is 1.8, \ie similar to that for the broad
components.

During the longest evolutionary phase of a supermassive binary, the system will
be in a quiescent state and the black hole separation is almost constant.
However, if the source undergoes some fuelling during this state, it will
become a strong non-thermal emitter (Begelman \etal 1980). In Section 3.2 we
showed that the optical to X-ray continuum of \rxj\ appears to be dominated by
a non-thermal component, thus this AGN may contain a supermassive binary which
is in a `high state', \ie in a fuelling phase.  In a system where one of the
components dominates, this activity would be concentrated at the larger hole.
In \rxj, the \Ha\ peak displacements suggest that the  components are similar
in size, so perhaps the activity is more evenly divided, producing the observed
line emission.  If low velocity gas has been ejected by the binary, then the
profiles may well have the double-peaked profiles observed. Meanwhile, each
very broad line region would see an additional ionizing continuum, perhaps
illuminating regions which would otherwise be in shadow and producing the
strong, broad Balmer lines. 

We conclude that the data presented here are consistent with the binary black
hole model for the double-peaked emission, although more data would be required
for a thorough evaluation, for example long-timescale variability studies could
be used to search for evidence of shifts in the positions of the peaks. The
orbital period of the binary may be estimated using the relationship
$P=550R_{\rm 18}^{3/2}M_9^{-1/2}$ yr (from Kepler's third law; Halpern \&
Filippenko 1988), where $R_{18}$ is the distance between the two components in
units of 10$^{18}$ cm and $M_9$ is the sum of the black hole masses in units of
10$^9 M_\odot$. Assuming that the binary separation is 1~pc (which is roughly
the distance to the outermost regions of the BLR in Seyfert galaxies) and that
the sum of the black hole masses is 10$^8~M_\odot$, (\ie from the Eddington
limit, assuming that the system is radiating at 0.2$L_{\rm Edd}$; see Section
3.2), this gives an orbital period of about a thousand years. Thus we would not
expect to see any changes in the peaks' positions over many years.
Alternatively, imaging data might reveal evidence of galaxy mergers, such as
regions of enhanced star formation like that seen in OX~169 (Stockton \&
Farnham 1991), bridges and tails.

\subsubsection{Double jets or cones}

In the jet model (\eg Foltz, Wilkes \& Peterson 1983; Zheng \etal 1990,1991),
material moving radially outwards is illuminated in a double-cone geometry. The
material itself may be emitted in a double-cone, or material moving out \eg
spherically may be lit up by a cone-shaped or otherwise anisotropic radiation
pattern. This model has many free parameters, providing the flexibility to fit
many different profile shapes and variability patterns (Zheng \etal 1990).
Zheng \etal\ (1991) fitted a double-stream model to the \Ha\ profile of
3C~390.3, consisting of a broad component with an opening angle of 60$^\circ$
and a narrower component with  an opening angle of 25$^\circ$; the latter
contributed 20 per cent of the total \Ha\ emission. The broad component
produces a flat profile while the peaks are emitted by the narrow component.

The second fit to the \Ha\ profile of \rxj\ (model 2 in Table 3) suggests a
crude analogy with this model, where the peaks are superimposed on one broad
underlying component. It is also possible that only the peaks are produced in
jets whereas the broad underlying component is emitted from a more typical BLR,
(which is thought to be disc-shaped for the low-ionization lines, \eg Krolik
\etal 1991; O'Brien, Goad \& Gondhalekar 1994). However the peaks in \rxj\ 
have smaller velocities than those inferred for 3C~390.3, suggesting that for
\rxj, perhaps {\sl (1)}  the jets extend to larger distances; {\sl (2)} the
jets are viewed at a greater angle to the line of sight; or {\sl (3)} the net
velocity of the material in the jets of \rxj\ is small. Note that for case (2)
however, if a dusty, molecular torus lies beyond the BLR of \rxj\ (see \eg
Antonucci 1993  and references therein), then this will occult much of the
receding jet when the system is viewed at large angles to the line of sight
(assuming that the torus' and jet's axes are coaligned) so that the red peak
would be very weak. 

In Section 3.2, we showed that the data imply an enhanced power-law component
common to the optical and X-ray ranges. We can speculate that this may be
related to a double-sided jet, \ie perhaps the jet was formed during an
outburst or a high state of non-thermal activity.  The extra ionizing continuum
might also power the very broad Balmer line emission (in the `normal'
high-velocity BLR). This concept would provide a link between an enhanced very
broad line flux, double peaks and a strong non-thermal continuum. It is 
similar to an idea previously suggested for NGC~1097 by Storchi-Bergmann \etal
(1993), \ie that the recent appearance of double-peaked Balmer emission in this
source was due to the formation of a new optical jet.

With so many free parameters for the jet model (\eg the opening angle of the
cone, inner and outer extents of the line-emitting portion, line emissivity and
velocity laws throughout the cone etc.; these may also be independent
parameters for the separate red and blue cones), constraints from the present
\rxj\ data are hard to find. If changes in the line emission from the jets are
induced by the central ionizing source, then the response in the red component
should lag the blue (and probably by several years; Livio \& Pringle 1996),
thus variability studies may provide an observational test of this model. For
the present however, we find that the \rxj\ data are consistent with a
double-sided jet model for the BLR.

\subsection{Double-peaked lines in radio-quiet AGN}

The rarity of double-peaked lines amongst radio-quiet objects seems puzzling.
Eracleous \& Halpern (1994) proposed that ADs in radio-quiet AGN  are
intrinsically different due to different accretion rates and the availability
of fuel. They suggest that an ion torus forms in the centre of radio-loud
objects  because the accretion rate is very low (this idea was originally
proposed by  Rees \etal 1982). The ion torus collimates the radio jets  and
provides a source of illumination for the outer edges of the AD, producing
double-peaked emission lines. In radio-quiet AGN, the accretion rates 
are generally higher so that the ion torus, and thus also the radio jets and
the double-peaked lines, would not form.

We have found that an AD is unlikely to be the source of the double-peaked
lines in \rxj\ and similar conclusions have been  reached for three other
double-peaked radio-quiet Seyferts (Marziani \etal 1992; Bower \etal 1996). So
perhaps the source of the double-peaked emission in radio-quiet AGN is
physically different from that in radio-loud objects, \ie if the AD model is
(in most cases at least) appropriate for radio-loud AGN, then a binary BLR or
double-jet may be more likely for radio-quiet objects. This in turn implies
that  supermassive binaries (where the black hole masses are similar in size
and nature) and bipolar line-emitting jets are both very rare. 

The binary BLR and bipolar jet models are also viable candidates for radio-loud
double-peaked AGN (as well as radio-quiet); why should these prefer a
radio-loud host? One obvious physical link between the supermassive binary and
radio-loud AGN models is that both are expected to have low accretion rates
(Begelman \etal 1980; Rees \etal 1982). In the bipolar jet model, the presence
of a biconical, outflowing BLR seems natural in an environment which must
already be dominated by the outflows of the radio jets.

\section{Conclusions}

\rxj\ is an X-ray selected AGN whose \Ha\ and \Hb\ emission lines exhibit
double-peaked profiles. The `blue' peak has a velocity of --1600 km s$^{-1}$
relative to the systemic velocity and the `red' peak has a relative velocity of
1000~\kms. Using combinations of Gaussians, we show that the  \Ha\ profile may
be modeled as {\sl (1)} two narrow peaks superimposed on one broad underlying
feature or {\sl (2)} as a pair of  lines each composed of a broad and a narrow
component, with the central wavelength of the broad component fixed to that of
its associated narrow line.

The {\sl ROSAT} PSPC spectrum is well-fitted by a power-law with an energy
spectral index, \ax=1.2$\pm$0.2 and shows no evidence for significant soft
X-ray absorption by cold gas or a strong soft X-ray excess. The optical
continuum has a similar slope (\aopt=1.0$\pm$0.5) with little evidence for a
strong BBB, except perhaps at the blue extreme of the spectrum (\ie at rest
wavelengths below $\sim$4000~\AA, although this may be the `little' blue bump,
a false continuum made of Balmer continuum emission and blends of \feii). It is
consistent with an extrapolation of the X-ray power-law and suggests that the
optical to X-ray continuum may be dominated by a non-thermal component. We
estimate that any possible thermal component contributes at most $\sim$40 per
cent of the total luminosity between log$\nu=$14 to 18~Hz.

We have compared these results with  three different models of the
double-peaked profiles, an AD, a binary BLR and a double-cone geometry of the
BLR. Both \Ha\ and \Hb\ show a net blueshift with respect to the \oiiiab\
lines, thus we find that AD models of the Balmer line emission, which predict a
net gravitational redshift, are not appropriate for this object. The binary BLR
model predicts that the two black holes must be similar in size and nature. The 
system would probably be in a `high' state, \ie accreting gas at an unusually
high rate, which may produce the strong non-thermal component which is
observed. The double-jet model is also possible; the data imply that
the velocity in the jets must be quite low which might be because the velocity
is intrinsically low, because of orientation effects or flows which reach to
relatively large distances.

\section*{Acknowledgments}

We are very grateful to Jim Condon for providing the VLA Sky Survey data. The
optical spectrum was provided by the RIXOS consortium and we thank all those
who have contributed to the RIXOS project. This research has made use of LEDAS,
the Leicester Database and Archive Service.

\bye